# Defect-Driven Anomalous Transport in Fast-Ion Conducting Solid Electrolytes

Andrey D. Poletayev[1,2*], James A. Dawson[3,4], M. Saiful Islam[5], Aaron M. Lindenberg[1,2*]

## Abstract

Solid-state ionic conduction is a key enabler of electrochemical energy storage and conversion. The mechanistic connections between material processing, defect chemistry, transport dynamics, and practical performance are of considerable importance, but remain incomplete. Here, inspired by studies of fluids and biophysical systems, we re-examine anomalous diffusion in the iconic two-dimensional fast-ion conductors, the β- and β``-aluminas. Using large-scale simulations, we reproduce the frequency dependence of alternating-current ionic conductivity data. We show how the distribution of charge-compensating defects, modulated by processing, drives static and dynamic disorder, which lead to persistent sub-diffusive ion transport at macroscopic timescales. We deconvolute the effects of repulsions between mobile ions, the attraction between the mobile ions and charge-compensating defects, and geometric crowding on ionic conductivity. Our quantitative framework based on these model solid electrolytes connects their atomistic defect chemistry to macroscopic performance with minimal assumptions and enables mechanism-driven 'atoms-to-device' optimization of fast-ion conductors.

## Introduction

The transport of charged ions in solid electrolytes underpins critical technologies, such as rechargeable batteries, fuel cells, and electrocatalysts, which are vital to the transition to low-carbon energy systems[1–3]. In the solid state, ionic diffusion and conductivity are mediated by thermally-activated rapid translations or hops of mobile ions between stable lattice sites. Hops are separated by relatively long residence times, and each ion executes a random walk. Based on Fickian diffusion, in the long-time regime of such a random walk, the time-averaged mean-square displacement (tMSD) of a mobile ion is linear with time[4]. The prediction of practical performance in the low-frequency, direct-current (DC) limit from atomistic simulations hinges on the key assumption of Fickian diffusion.

The prediction of macroscopic ionic conductivity and of its dependence on material processing[5–7] in turn enables device design[8]. The frequency-dependent conductivity is described by the net movement of charge[9], i.e., the ensemble-averaged mean-square displacement (eMSD) of all ions. At picosecond to nanosecond timescales accessible to atomistic simulation, slowly-evolving dynamics such as the fluctuating positions of other mobile ions and the motions of the host structure[10–16] may couple to the transport of hopping ions. Phenomenological mesoscale models incorporate disorder via randomly frozen energetic barriers[17], dynamic relaxation[18], or nonlinear terms in the diffusion equation[19]. However, a general frequency dispersion in conductivity highlighted by Jonscher[20] complicates the prediction of practical macroscopic ionic conductivities from atomistic simulation.

[1] Stanford Institute for Materials and Energy Sciences, SLAC National Laboratory, Menlo Park, CA, USA
[2] Department of Materials Science and Engineering, Stanford University, Stanford, CA, USA
[3] Chemistry – School of Natural and Environmental Sciences, Newcastle University, Newcastle upon Tyne, UK
[4] Centre for Energy, Newcastle University, Newcastle upon Tyne, UK
[5] Department of Chemistry, University of Bath, Bath, UK
* Correspondence to andrey.poletayev@gmail.com, aaronl@stanford.edu

Inspired by studies of anomalous transport in fluids[21] and biophysical systems[22–26], we re-examine ion conduction in the iconic yet complex families of model solid electrolytes, the β- and β``-aluminas. In these materials, an unusual frequency dispersion of conductivity has been measured[27–29] at timescales as long as microseconds, and corrosion kinetics consistent with sub-diffusion (which is defined as a sub-linear dependence of mean-square displacement of mobile ions on time) have been observed[30], but the mechanisms of both phenomena have yet to be explained.

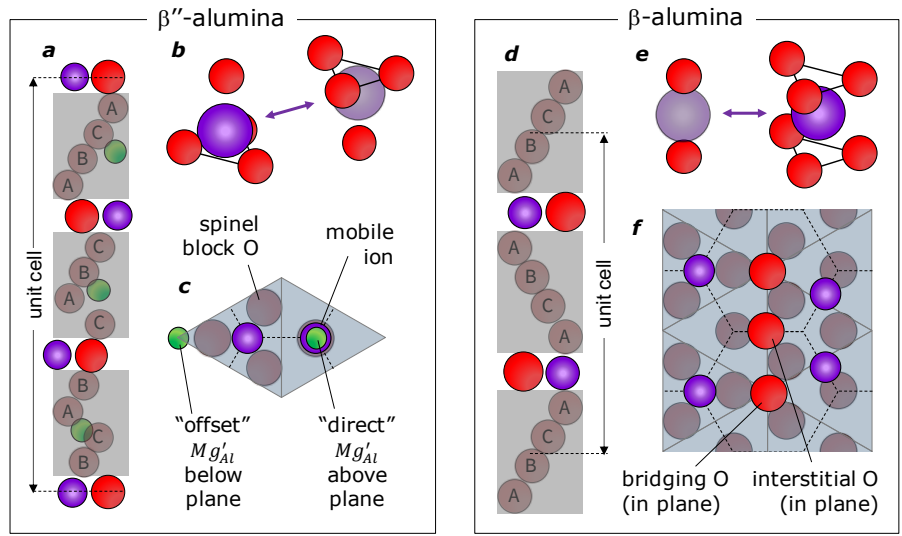

**Figure 1 | Crystal structures, conduction-plane sites, and charge-compensating defects in β- and β''-aluminas. a**, In the β''-alumina structure, conduction planes are sandwiched between blocks of spinel $\gamma$-alumina (grey). **b**, Mobile ions (purple) hop between identical four-coordinate sites. **c**, Charge-compensating $Mg_{Al}$ defects in β''-alumina locate ≈6 Å above and below the conduction planes within the spinel blocks. The β-alumina structure (**d**) differs from the β'' phase by the stacking sequence of close-packed oxygen layers (ABCA), which are symmetric about the conduction planes. **e**, In β-aluminas, mobile ions hop between low-energy six-coordinate Beevers-Ross sites and high-energy two-coordinate anti-Beevers-Ross sites. **f**, Charge-compensating interstitial oxygen ions in β-aluminas locate at the boundaries (edges) of conduction-plane sites and form long-lived clusters with four mobile ions.

The β- and β``-alumina materials (Figure 1) exhibit fast ionic conduction of alkali metal ions $M^+$ (M = Na, Ag, or K) within two-dimensional planes having honeycomb site connectivity. The β`` structure has one type of four-coordinate site (Figure 1b). Two temperature regimes of activation[31,32] are attributed to intermediate-range ordering[33,34] of mobile ions near room temperature. However, ionic conductivity has been modulated by quenching via re-distributing charge-compensating $Mg_{Al}$ defects[35,36]. By contrast, in β-aluminas ions hop between two non-equivalent sites (Figure 1e)[13,37,38], and the activation energy is constant at all temperatures. Charge-compensating oxygen interstitials bind mobile ions (Figure 1f)[39,40] and block diffusion paths. Processing is known to modulate ionic conductivity[41], but no complete mechanism is known.

Using long-time molecular dynamics (MD) simulations with long-established interatomic potentials (Methods), we focus on the dynamics of hopping ions in the dispersive regime (ps-ns) that connects the DC regime of practical application to the vibrational frequencies. The simulated stoichiometries are $M_{1.2}Al_{11}O_{17.1}$ for β-aluminas, and $M_{5/3}Al_{31/3}Mg_{2/3}O_{17}$ for β``-aluminas, for M = Na, Ag, or K, and simulation length $\Delta = 100$ nanoseconds. Such potentials-based atomistic techniques have the major advantage of examining ion conduction processes at length- and timescales that are at least two orders of magnitude greater than *ab initio* MD methods.

Here we demonstrate a computational approach that quantitatively connects atomistic defect chemistry to macroscopic conductivity in these model ionic conductors. First, we characterize persistent sub-diffusion and non-Gaussian statistics. Second, we quantify deviations from ergodicity, which complicate the prediction of a practical DC conductivity. Hence, we calculate the alternating-current (AC) ionic conductivity in excellent agreement with literature spectra. We then explain the anomalous transport and the frequency dispersion of AC conductivity by analyzing the nanometer-scale defect chemistry of β/β``-aluminas. Finally, we compare distributions of defects that correspond to variations in processing conditions. Together, our mechanistic study shows how defect chemistry and disorder determine the energy landscape and practical conductivity in solid electrolytes.

## Statistical Descriptors of Ion Transport

We consider first the mean-square displacements of mobile ions, time-averaged over trajectories (tMSD, Figure 2a) for Na β``-alumina. From short to long timescales $t$, an ion sequentially undergoes ballistic ($t^2$), bound ($t^0$), and diffusive dynamics. However, the expected Fickian dependence ($t^1$) is only reached at elevated temperature. At 300 K and below, the simulated transport is sub-diffusive, characterized by an exponent below unity even for long timescales (≥10 ns, Figure 2a) and average displacements significantly larger than one unit cell (Figure 2b).

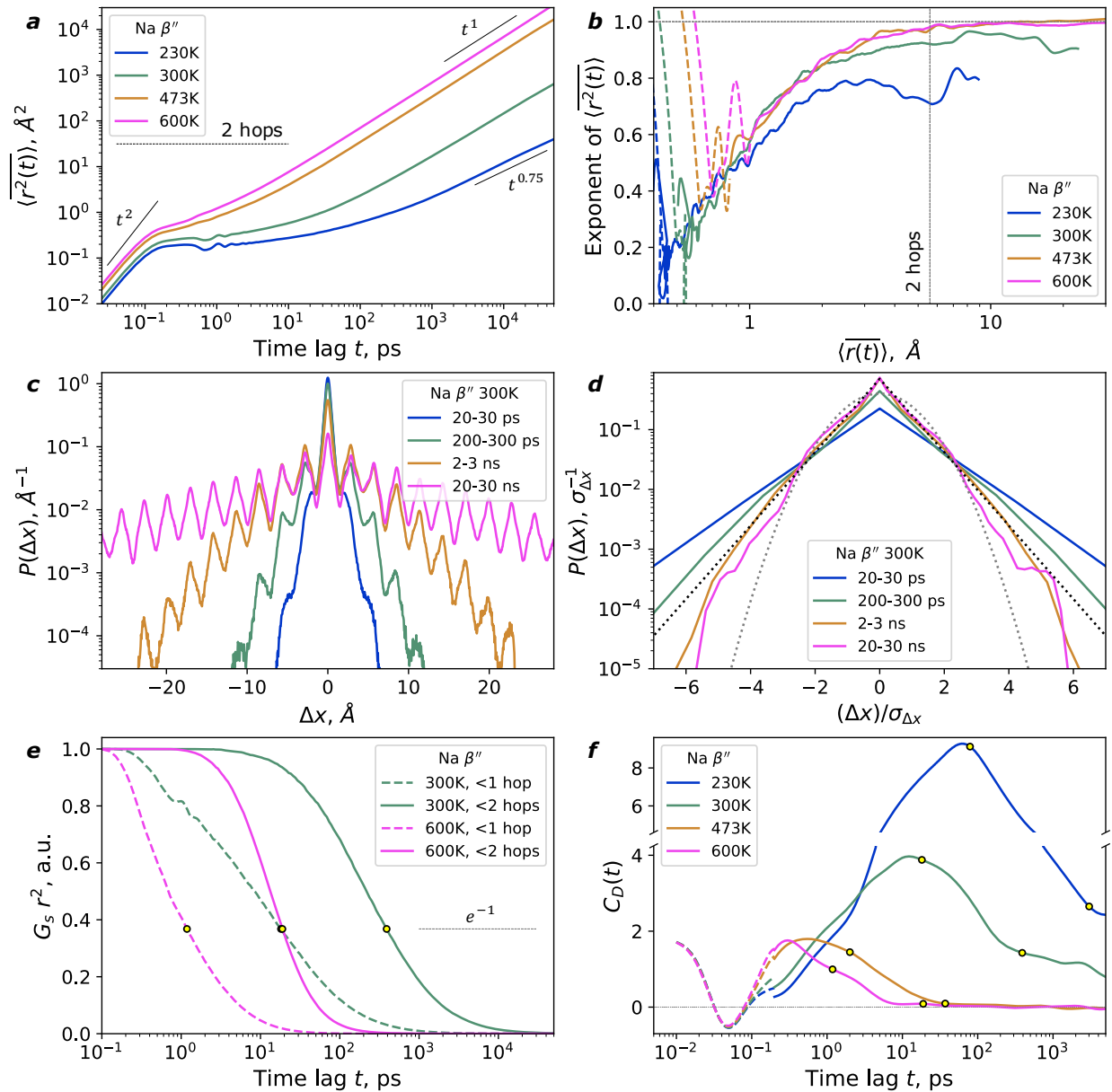

**Figure 2 | Ion diffusion in Na β``-alumina. a**, Time-averaged mean-square displacement tMSD of mobile ions. **b**, Exponent of tMSD vs. time-averaged displacement of ions. For the same average displacements, the exponent at ≤300 K is lower than at ≥473 K. In **b**, the horizontal guide is the Fickian limit $t^1$, and the vertical line is one unit cell (5.6 Å, 2 hops). **c**, Time slices of the distribution of ion displacements $\Delta x$ along [100] at 300 K. **d**, Distributions of ion displacements $\Delta x$ along [100], each rescaled by its standard deviation $\sigma_{\Delta x}$. Laplace and Gaussian distributions are shown as black and grey dotted lines, respectively. **e**, The probability of an ion remaining within 1.7 Å (<1 hop, dashed) or 4.6 Å (<2 hops, solid) of an initial position after a time lag $t$. The relative change in the timescale due to varying the distance cutoff by 0.1 Å is ≤10%. **f**, Diffusion kernel correlation $C_D$, with one- and two-hop relaxation times as derived from (**e**) shown as yellow circles. In **b** and **f**, short-time checks of the ballistic limits (dashed), tMSD $\propto t^2$ and $C_D \to 2$, respectively, used short, 250-ps trajectories. Corresponding metrics for other simulated materials are shown in Extended Data Figures 1-6.

The non-Fickian long-time behavior of the average tMSD at 300 K motivates an examination of the full distributions of ion displacements[22,25]. Figure 2c shows such distributions along [100] in Na β``-alumina at 300 K, with obvious peaks at lattice sites every 2.8 Å. The same distributions, aggregated to one point per 2.8-angstrom lattice site, and each rescaled by its standard deviation, are shown in Figure 2d. At short timescales, the distributions are wider than a Laplace distribution, and even for 20–30 ns remain wider than a Gaussian. While short-time deviation is expected given the activated nature of ion hopping, the persistence of a non-Gaussian distribution at long time lags suggests deviations from normal diffusion impact the macroscopic conductivity. Of all the simulated materials, only K β``-alumina reaches a Gaussian distribution at 300 K during the simulation.

We begin to quantify hopping kinetics using the self part of the van Hove function, the Fourier counterpart of the experimentally observable incoherent intermediate scattering function for the mobile ions[42]. The decay of the probability of an ion remaining within 1.7 Å (<1 hop) or 4.6 Å (<2 hops) of an initial position (Figure 2e) provides a simple proxy for hopping timescales. The probability follows Mittag-Leffler behavior: stretched-exponential at short timescales, followed by power-law asymptotic long-time behavior. For simplicity, we use 1/e decay times (Figure 2e, yellow circles) as proxies for the timescales for one and two diffusion events. We refer to them below as "one-hop relaxation time" and "two-hop relaxation time" by analogy with the structural relaxation interpretation of the van Hove function[43].

## Deviations from Ergodicity

The anomalous features of the statistics in Figure 1a-d motivates a quantitative exploration of the distributions beyond their averages, i.e., dynamic heterogeneity. We compare three descriptors of dynamic heterogeneity: the diffusion kernel correlation ($C_D$, Figure 2f)[21], the non-Gaussian parameter (NGP, Figure S1), and the relative variance of the average tMSD as the ergodicity breaking parameter ($EB$, Extended Data Figure 7)[22,44].

Within continuous-time random walk models, the diffusion kernel yields the second moment of displacement[45] that determines the tracer diffusion coefficient. Its correlation $C_D$ characterizes the coupling of environmental dynamics to diffusion[21]. Two timescales are of interest. First, the times when $C_D$ peaks are interpretable as the timescale of fluctuation modes that mediate diffusion. In Na β``-alumina, $C_D$ peaks close to the 1-hop relaxation time at ≤300 K, but at a shorter time at ≥473 K (Figure 2f). By contrast, in all β-aluminas, $C_D$ peaks after the one-hop relaxation time, and the 2-hop relaxation begins a long-time regime when $C_D$ slowly decreases (Extended Data Figures 1,4,6).

Second, the timescale at which $C_D$ reaches an asymptotic value represents the onset of a long-time regime where average values may suffice to characterize diffusion. At ≥473 K for Na β``-alumina, the two-hop relaxation time comes already after $C_D$ asymptotically approaches zero ($C_D \rightarrow 0$). By contrast, for Na β-alumina $C_D$ approaches zero only at 1000 K, and only at $t \geq 1$ ns. For Na β``-alumina, the NGP (Supplementary Note 1), like $C_D$, exhibits two temperature regimes at long timescales. The low-temperature regime is consistent with persistent heterogeneity and sub-diffusive transport, and is similar to that in β-aluminas, while the high-temperature regime remains distinct.

This difference highlights the distinct diffusion mechanisms between the β`` and β structures. Because of its two non-equivalent mobile-ion sites (Figure 1e), single hops in Na β-alumina create dynamic heterogeneity, and only double hops constitute diffusion events at all temperatures. In Na β``-alumina, with one type of mobile-ion site (Figure 1b), at high temperatures single hops suffice for diffusion, but the low-temperature regime also requires diffusion by double hops. Notably, this low-temperature regime is absent in K β``-alumina (Extended Data Figure 2). We investigate the origin of the low-temperature regime in Na β``-alumina and of its absence in K β``-alumina below.

Finally, the relative variance in the average tMSD for an arbitrary short timescale $t$ ("ergodicity breaking" parameter $EB$) should collapse with the inverse of the overall length of simulation (Δ) for the ergodic assumption to hold[21–23,44]. However, this does not hold at ≤300 K for any simulated material, and is only reached at elevated temperatures (Extended Data Figure 7). Thus, at room temperature, the diffusion of mobile ions in both Na β- and β``-alumina remains weakly non-ergodic within Δ = 100 ns, and the trajectories of individual ions remain distinguishable to macroscopic timescales. This is consistent with experimentally observed non-Debye dispersion of AC conductivity in large single crystals[27–29]. For practical applications, our key findings of sub-diffusion and ergodicity-breaking imply that singular, constant-valued transport coefficients in the DC limit at 300 K cannot be derived.

## Ionic Conductivity

Due to non-Fickian diffusion and deviations from ergodicity in our simulations, we adopt a stochastic diffusion framework to calculate the AC ionic conductivities directly from the mean-square charge displacement via the eMSD The values of eMSD sampled from the simulated 100-nanosecond trajectories for any given time lag $t$ in the ps-ns range are distributed exponentially, with one parameter sufficient to describe both the mean and variance at each $t$ (Extended Data Figure 8). We compare the conductivities calculated using the means of eMSD distributions with literature experimental AC conductivity spectra of β/β``-aluminas at frequencies ν=1/$t$ (Figure 3a-c)[27–29,41,46,47]. Our simulations reproduce the Jonscher regime with conductivity $\sigma \propto \nu^{0.7}$ for Na β``-alumina at 300 K (Figure 3a) or $\sigma \propto \nu^{0.6}$ for Na and Ag β-aluminas at 300 K (Figure 3bc). More importantly, our simulations reproduce the weaker dispersion $\sigma \propto \nu^{0.1}$ that extends to lower frequencies at room temperature, in agreement with impedance measurements of centimeter-size single crystals[27–29] (Figure 3a-c). In our simulations, this additional dispersion disappears at elevated temperature, concurrent with the disappearance of anomalous diffusion. This suggests that long-time dispersion extending to macroscopic timescales is a consequence of the anomalous diffusion.

For the longest time lag used, $t$ = 2.5 ns, we also compare our results to previous experimental measurements in the low-frequency limit[31,32,36,48–50] (Figure 3d-f) while acknowledging that at low temperatures the simulated conductivities overestimate experimental ones because of the remaining frequency dispersion. The agreement of simulated AC and low-frequency conductivities with experiment for Na and K β``-aluminas is better than within a factor of 2 at ≥300 K. This provides further validity that our simulations successfully model the macroscopic performance of β``-aluminas at room temperature.

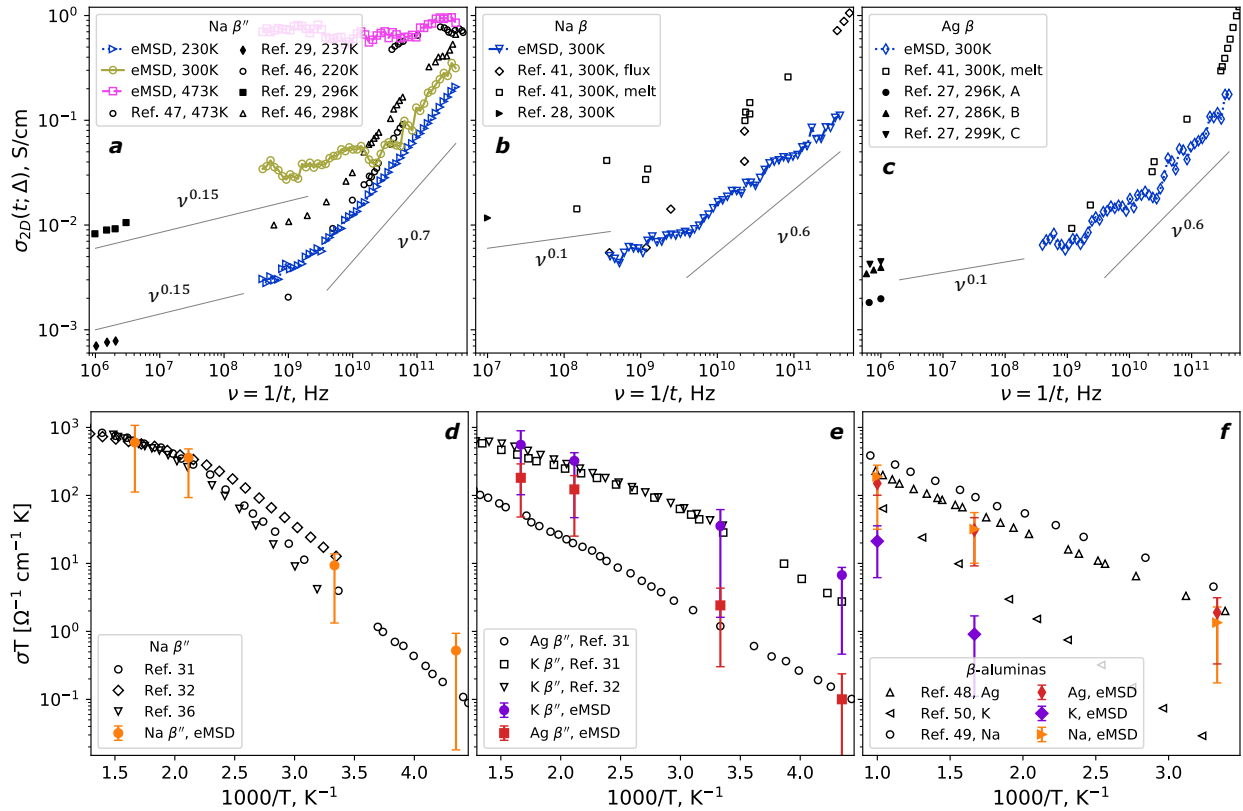

**Figure 3 | In-plane ionic conductivity in β/β``-aluminas: Simulated AC frequency spectra (a-c) and Arrhenius relations (d-f) and comparison to experiment.** Spectra: **a**, Na β``-alumina at 230 K, 300 K, and 473 K. **b**, Na β-alumina at 300 K. **c**, Ag β-alumina at 300 K. Arrhenius relations: **d**, Na β``-alumina, **e**, K and Ag β``-aluminas, **f** Ag, Na, and K β-aluminas. Filled symbols in (**a-c**): literature measurements of electrical impedance. In (**d-f**), $t$ = 2.5 ns (400 MHz), and error bars denote 20th–80th percentiles of the sampled eMSD. Power-law scaling relations (grey lines in **a-c**) are guides to the eye, not fits.

For β-aluminas, except for Ag, simulated conductivities underestimate experiments (Figure 3bcf). However, the discrepancy propagates from the fastest, vibrational timescales and does not arise from long-time behavior discussed here. The good agreement for Ag β-alumina (Figure 3c) likely comes from using two-coordinate Ag, e.g. in $Ag_2O$, as a model for the Ag-O interaction; no natural analogs for Na/K exist. The ability of Ag to effectively shrink in a two-coordinate configuration underpins fast ionic conduction in Ag β-alumina, while Ag β``-alumina is instead the slowest of its family. Our simulation likely underestimates the success rates of hops into the high-energy anti-Beevers-Ross (aBR) sites in Na and K β-aluminas. For K β-alumina, the aBR site constitutes the most stringent bottleneck, and our simulation underestimates conductivity most strongly (Figure 3f). The absence of this structural bottleneck in K β``-alumina makes it the fastest room-temperature ionic conductor of the six simulated materials, and accounts for the agreement between simulation and experiment (Figure 3e).

## Nanoscale Defect Chemistry and Static Disorder

Next, we use the large pool of diffusion events from our long-time simulations to quantify the influence of defect chemistry on ion transport. The charge-compensating defects provide a natural coordinate system to map the hops of mobile ions within the conduction planes. Each diffusion event, or non-returning hop using the Funke definition[18], has an origin site and a destination site. In β``-aluminas, most conduction-plane sites are in proximity of multiple defects, and the number of such neighbors is the relevant metric. Additionally, we distinguish the relative positions of $Mg_{Al}$ defects in the spinel block: the defect can either sit "directly" above/below the conduction-plane site, or "offset", above/below one of the site's corners (Figure 4a). Figure 4b-d shows the rates of diffusion for each pair of neighboring sites, aggregated by the number of Mg defect neighbors at the origin site (horizontal axis) and destination site (vertical axis). In Na and Ag β``-aluminas, sites with fewer neighboring defects exhibit faster rates of diffusion events, while in K β``-alumina the opposite holds.

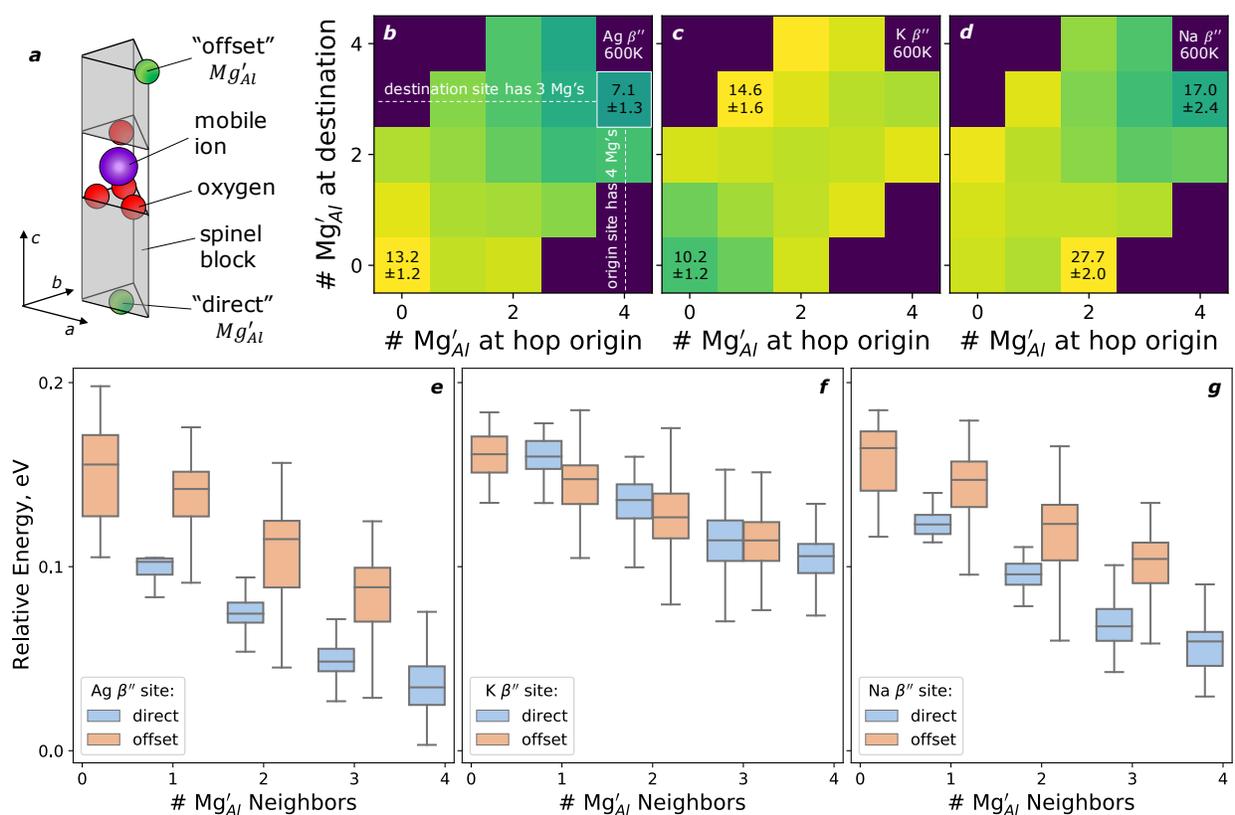

**Figure 4 | Site energetics in β``-aluminas. a**, Schematic of two types of $Mg_{Al}$ defect positions relative to conduction-plane sites: directly in the center (bottom) and offset to a corner (top). The example conduction-plane site shown in (**a**) is counted as having two Mg neighbors. Both distances are ≈6 Å. **b-d**, Rates of diffusion, in units of hops per edge and per nanosecond, classified by their sites of origin (horizontal axes) and destination (vertical axes), in Ag, K and Na β``-aluminas respectively at 600 K. All color scales are normalized to the noted maximum rates (yellow). For example, the highlighted square in (**b**) corresponds to a rate of 7.1±1.3 diffusion events per edge per nanosecond from sites located next to four Mg defects to sites located next to three Mg defects. The uncertainty value is a standard error over the edges that fit this criterion. **e-g**, Distributions of

relative Helmholtz free energies for all conduction-plane sites in β``-aluminas at 600 K, when ergodicity is shown to hold, aggregated by whether each site has one "direct" Mg defect neighbor (blue) or none (orange) as defined in (**a**).

When ergodicity holds, the time-averaged occupancy for each conduction-plane site corresponds to relative site energies and quantifies static energetic disorder among sites. Figure 4e-g shows relative Helmholtz free energies for mobile ions in all conduction-plane sites extracted from time-averaged site occupancies. As expected from the dopant $Mg_{Al}$ carrying one net-negative charge relative to the lattice, sites with more defect neighbors are attractors of mobile ions and are on average more fully occupied (with lower average Helmholtz free energies by ≈25–30 meV/neighbor).

The more interesting comparison is between sites of different types ("direct" or "offset") for an equal total number of Mg defect neighbors. In Ag and Na β``-aluminas, the sites with a "direct" neighboring defect (Figure 4e-g, blue) are more attractive than sites without them (Figure 4e-g, orange), by ≈60 and ≈40 meV, respectively. This energetic difference causes "offset" sites to anchor the mobile-ion vacancy ordering at low temperatures (Supplementary Note 2), and is responsible for the increased low-temperature activation energy. In K β``-alumina, the sites with and without "direct" $Mg_{Al}$ defect neighbors are energetically indistinguishable, and the activation energy remains almost constant with temperature (Figure 3e).

The site-specific diffusion rates and free energies for β-aluminas, corresponding to Figure 4, calculated for 1000 K (Figure S6), show average free energy differences between BR and aBR sites higher than the simulated activation energy for conduction, as in previous works[38]. This result suggests that an interstitialcy-like knock-on mechanism where the repulsion between mobile ions transiently comprising an interstitial pair on neighboring BR and aBR sites lowers the energy barrier for hopping.

## Defects Drive Heterogeneity in Transport Dynamics

Static energetics predict vacancy ordering, but do not explain why diffusion proceeds via distinct sites among the three β``-aluminas (Figure 4b-d). Hence, we quantify the location-dependent kinetics of diffusion events. The memory of a hop's origin is characterized by the residence-time-dependent deviation from unity of the correlation factor $f = (1 - \langle cos\theta \rangle)/(1 + \langle cos\theta \rangle)$, where θ is the angle between consecutive hops. In β``-aluminas, $f$ is near-zero at short residence times, and increases for longer residence times towards the random value of 1 (Figure 5). At 300 K, the timescales for this randomization are longer than two-hop relaxation times (Figure 5, yellow circles). Instead, the timescale of $C_D \to 0$ is the best descriptor of the kinetics of memory loss. This is consistent with the derivation of $C_D$ from the evolution of a random-walker's environment[21,45], and with Funke's dynamic relaxation model[18]. At 300 K, all diffusion events are non-random. This is consistent with a simple mechanism for dynamic relaxation: an ion randomizes following a hop only as fast as other ions also hop around it. Additionally, a waiting-time memory is present, as for a continuous-time random walk, at short residence times (Supplementary Note 3).

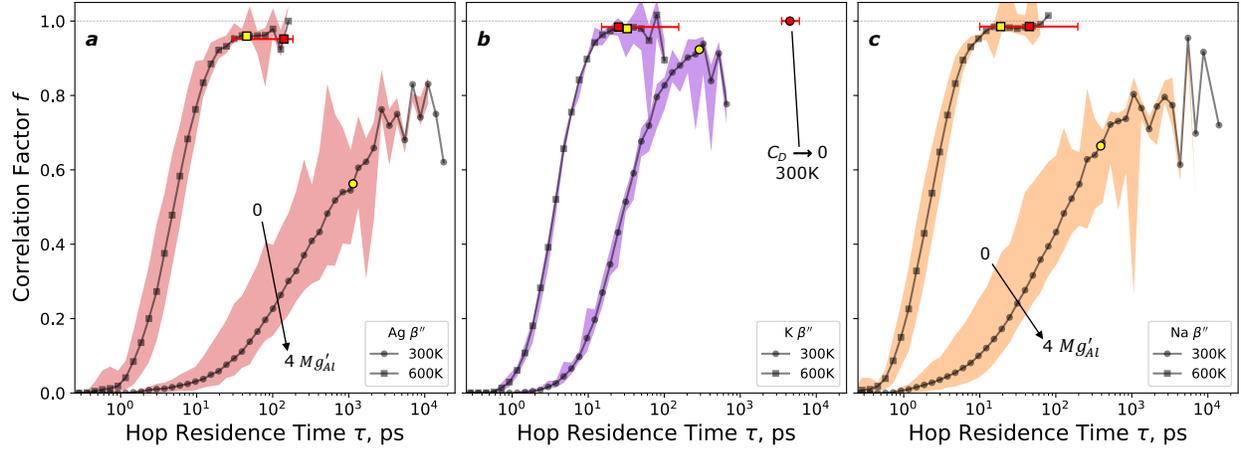

**Figure 5 | Correlation factor in β''-aluminas as a function of hopping residence times and locations: (a) Ag, (b) K, and (c) Na.** Lines: correlation factor, averaged over hops to all sites. Circles: 300 K, squares: 600 K. Shaded regions: ranges of location-dependent correlation factors, aggregated using the same mapping metric as in Figure 4. The sites with the most neighboring defects always have the lowest correlation factor for a given residence time (Extended Data Figure 9a-c). Yellow symbols: two-hop relaxation time from $G_s$ (square: 600 K, circle: 300 K). Red horizontal ranges: approximate timescales for $C_D \approx 0$ at 600 K. At 300 K, all hops are correlated: only for K β''-alumina does $C_D \to 0$ (≈4-5 ns) within the 100-ns simulations.

The shading around the average curves in Figure 5 shows the location dependence of $f$, or relaxation kinetics for sites with different numbers of neighboring $Mg_{Al}$ defects. This defect-induced heterogeneity of relaxation kinetics is largest for Ag (wide shaded regions in Figure 5a), smaller for Na (Figure 5c), and negligible for K (Figure 5b). For Na and Ag β''-aluminas, hops to sites with fewest neighboring defects are fastest to randomize (Figure 5a,c, top/left edges of the shaded regions), explaining the fact that these emptier sites contribute most to diffusion (Figure 4b,d). By contrast, slower relaxation at sites with more neighboring defects (Figure 5a,c, bottom/right edges of the shaded regions) impedes diffusion at those sites. The overall diffusivity of ions in Na and Ag β''-aluminas is location-dependent on the nanometer scale, determined by the distribution of charge-compensating defects, despite the defects being ≈6 Å away within rigid spinel blocks. By contrast, in K β''-alumina, the memory timescale is independent of location (the location-resolved curves in Extended Data Figure 9, corresponding to the shading in Figure 5b, coincide), and a site's contribution to diffusion tracks its time-average occupancy (Figure 4c,f). This is consistent with repulsions between the K ions dominating the attraction between K ions and static defects. We conclude that repulsion between mobile ions drives conductivity in K β''-alumina.

The location dependence of dynamic relaxation is exacerbated in β-aluminas (Extended Data Figure 9): the charge-compensating oxygen interstitial is doubly-charged and located immediately adjacent to mobile ions. The interstitial associates or binds mobile ions in a defect cluster[39,40]. In β-aluminas, the slow (≥10 ns at 600 K) exchange of mobile ions between free and defect-bound states determines the timescale of reaching ergodicity. We propose that this slow exchange and the resulting location dependence of ion mobility drive the dispersion of conductivity ($\sigma \propto \nu^{0.1}$ regime in Figure 3a-c) at 300K, the same way the ordering of mobile-ion vacancies does in β''-aluminas.

## Tuning Disorder and Conductivity with Processing

Finally, we consider tuning the relative positions of defects, as they may be engineered with processing[5–7,35,36]. Our simulation shows that sites without "direct" $Mg_{Al}$ neighbors anchor the ordering of mobile-ion vacancies in Na β''-alumina. Whereas in a slowly-cooled material the "direct"-neighbor sites belong to half of the six possible $\sqrt{3} \times \sqrt{3}$ conduction-plane sublattices, and vacancy ordering proceeds on the other sublattices, quenching distributes "direct"-neighbor sites randomly between the sublattices (Supplementary Note 2). This makes all sublattices equivalent, destabilizes the ordering of mobile-ion vacancies and increases room-temperature conductivity to match the high-temperature regime. The effect is absent in K β''-alumina (Figure S3), where repulsions between mobile ions control conductivity.

In the β-aluminas, mobile ions screen the Coulomb field of oxygen interstitials at the expense of forming defect clusters (Supplementary Note 4) that directly block diffusion pathways. The could further hinder diffusion by isolating a fraction of mobile ions in tortuous landscapes between the stationary clusters, and is termed 'geometric crowding'[22,24]. If present, crowding would be expected to manifest as increased dispersion when tMSD of mobile ions becomes comparable to the distances between defects. We simulate two quasi-random distributions of interstitials in Na β-alumina (Supplementary Note 5): located ≥1 site apart (more disordered), and located ≥4 sites apart (more ordered). The simulation with the higher disorder exhibits lower conductivities at long time lags, but not at short ones, and higher heterogeneity of dynamics (Figure S8). In our simulation, disorder among interstitials decreases conductivity by ≈50% at 1 ns. All features are consistent with increased geometric crowding when interstitials are disordered. Due to the different effects of the Coulomb field of charge-compensating defects in β''-aluminas vs β-aluminas (extended vacancy ordering vs formation of blocking clusters), processing that yields a kinetically controlled (quenched, disordered) distribution of defects improves the former, but thermodynamic control (ordered, well-separated defects) improves the latter.

## Discussion and Conclusions

Here, we have investigated how charge-compensating defects drive sub-diffusive transport dynamics in model solid-state ion conductors. Accounting for anomalous diffusion using a simple stochastic approach, we reproduce the frequency dispersion of AC conductivity in the Jonscher regime ($\sigma \propto \nu^{0.6}$), and at lower frequencies ($\sigma \propto \nu^{0.1}$). The latter is caused by the slow kinetics of exchange of mobile ions between lattice sites and immobilized within defect clusters (β-aluminas), or within a mobile-ion/vacancy superlattice (β''-aluminas). Our approach quantitatively predicts macroscopic ionic conductivity even with sub-diffusive dynamics.

The diffusion kernel correlation $C_D$ characterizes location-dependent dynamic relaxation that accompanies diffusion events (Figure 5). Comparing the timescales at which $C_D$ peaks with one-hop and two-hop relaxation timescales yields the mechanism of diffusion. In β''-aluminas, a switch from diffusion via double hops at the edges of ordered domains at low temperature to unimpeded diffusion via single hops at high temperature is consistent with the decrease of the activation energy upon heating. However, since the spatial distribution of charge-compensating defects anchor the local energetic disorder and room-temperature ordering of mobile ions, we conclude that the room-temperature sub-diffusive transport and the switch of dynamic regimes upon heating are both ultimately driven by defect chemistry.

In conclusion, using quenching in β``-aluminas[36] and geometric crowding by interstitials in β-aluminas as examples, we demonstrate new mechanistic insights into processing-disorder-performance relationships of solid electrolytes. In both cases, a redistribution of charge-compensating defects modulates practical ionic conductivity. Our approach highlights the mechanisms by which thermodynamic versus kinetic control of defect distributions during processing controls performance. In general, we present a quantitative framework for ionic conduction between the atomistic and macroscopic timescales, which is a key enabler of 'atoms-to-devices' multiscale modeling. Our approach offers pathways for the understanding and optimization of contemporary ion-conducting electrolytes for a wide range of energy-related applications.

## Acknowledgements

This work was supported by the Department of Energy, Office of Science, Basic Energy Sciences, Materials Sciences and Engineering Division, under Contract DE-AC02-76SF00515. MSI and JAD gratefully acknowledge the EPSRC Programme Grant "Enabling next generation lithium batteries" (EP/M009521/1). J.A.D. gratefully acknowledges the EPSRC (EP/V013130/1), Research England (Newcastle University Centre for Energy QR Strategic Priorities Fund) and Newcastle University (Newcastle Academic Track (NUAcT) Fellowship) for funding. Via membership of the UK's HEC Materials Chemistry Consortium, which is funded by the EPSRC (EP/L000202, EP/L000202/1, EP/R029431 and EP/T022213), this work used the ARCHER UK National Supercomputing Service. A.D.P. also thanks Geoff McConohy, Dr. Aditya Sood, Dr. Stephen Kang, and Philipp Muscher for helpful and invigorating discussions.

## Author Contributions

A.D.P. initiated the application of anomalous-transport concepts with advice and support from A.M.L. A.D.P. carried out simulations with instruction and help from J.A.D and advice from M.S.I. A.D.P. carried out analysis. A.M.L. advised and supervised the work. All authors contributed to the writing of the manuscript.

# Supplementary Information : Defect-Driven Anomalous Transport in Fast-Ion Conducting Solid Electrolytes

## Methods

Supplementary Note 1: Long-time Behavior of the Non-Gaussian Parameter
Supplementary Note 2: The Origin of Mobile-Ion Vacancy Ordering in β``-aluminas
Supplementary Note 3: Wide-tailed Distributions of Hop Residence Times
Supplementary Note 4: Site Energetics and Formation of Defect Clusters in β-aluminas
Supplementary Note 5: Geometric Crowding by Oxygen Interstitial Clusters
Supplementary Note 6: Trends with Mobile-Ion Size and Concentration

## Extended Data Figures 1-9

## Methods

The scripts and templates to generate simulation structures, and run simulations and analyses are available at https://github.com/apoletayev/anomalous_ion_conduction . The python packages *freud-analysis*[1] and *networkx*[2] were used extensively to generate the quasi-random structures. The perceptually uniform *batlow* color scheme[3] was adapted for many visualizations.

### Molecular Dynamics Simulations

Buckingham (exponent + 6$^{th}$ power) pairwise potentials were used with a long-range Coulombic cutoff at 12 Å. These were first developed for Na β-alumina[4] and used extensively since[5–9]. For K and Ag ions, the Binks library potentials were used[10]. All simulations were carried out using the LAMMPS simulation package[11] using a Nosé-Hoover thermostat. All ionic charges were integral, i.e. +1 for all mobile ions etc. Every simulation was started with the same random seed. The simulation volume was annealed at 1000 K for 20 ps using an NPT ensemble, cooled to simulation temperature over 6 ps, and simulated for an NVT run of length Δ = 100 nanoseconds with a timestep of 1 femtosecond to accumulate hopping events and average statistical descriptors for long time lags. The only simulation executed for Δ = 300 ns was Na β``-alumina at 230 K. The simulation of "quenched" K and Ag β``-aluminas were executed for Δ = 10 ns.

### Quasi-Random Placement of Charge-compensating Defects

Little is known about the precise locations of $O_i$ in β-aluminas relative to one another. Here we make an assumption of thermodynamic control over their relative distribution. Since they carry a -2 charge each, they should repel each other, but we are aware of no evidence for a defect superlattice in β-aluminas. We assumed that during crystal synthesis, typically multiple days above 1500 C and slow cooling[12,13], the $O_i$ defects are relatively mobile and settle to a thermodynamically favorable arrangement, i.e. far away from one another. For simulation, $O_i$ were placed quasi-randomly within the conduction planes starting with a defect-free material, with a minimum distance of 5 lattice sites (Beevers-Ross or anti-Beevers-Ross) between closest defects. This was the maximum distance attainable with an otherwise-random placements of defects. From 30 randomly-generated configurations of defects, one was picked with the lowest z-score for the distribution of numbers of sites versus distance from a nearest defect to ensure the structure could be considered representative for the given set of constraints. Similar algorithms were used to simulate a more random distribution of defects. A representative configuration was chosen with a less restrictive constraint, minimum distance of 2 sites between nearest defects. This example modification is one possible response to processing conditions in β-aluminas, but not the only one.

For β``-aluminas, Mg defects were substituted onto Al(2) sites according to experimental crystallographic studies[8,14]: only one half of Al(2) sites was used to approximate the "un-symmetric" distribution. Each Mg defect is "directly" above or below one conduction-plane site, and "offset" from 6 sites – but in a second conduction plane. Each conduction-plane site is affected by the defects on both sides of the conduction plane. Within the "un-symmetric" distribution of defects, any conduction-plane site has at most one "direct" Mg defect associated with it, and at most 3 "offset" ones. For comparison, a quenched distribution of defects (Figure S9) was simulated by placing the Mg substituents randomly on any Al(2) sites enforcing a minimum separation of 1 Al(2) between them as above for β-aluminas. Up to two "direct" defect neighbors are possible for a conduction-plane site in this case, one above and one below.

Multiple-Starts Sampling and Statistical Quantities

The non-Gaussian parameter and diffusion kernel correlation were computed as follows. The 100-nanosecond trajectory of each ion was subdivided into 50-nanosecond chunks starting at 0 simulation time, and spaced 70 ps apart, until 50 ns, approximately 700 starts for each ion. From all these, taken together for all mobile ions, the displacements Δx along [100] were averaged for given time lags *t* by combining the distributions of displacements within 20% of that time lag, i.e. 8-12 ps for *t*=10 ps, or 2-3 ps for *t*=2.5 ps (Figure 2c and equivalents). These distributions were binned in 2.8-Å bins corresponding to lattice sites (corresponding to the periodic fluctuations in Figure 2c) to yield a single number per bin. These binned distributions were rescaled by their standard deviations[15] to yield Figure 2d and equivalents. The self part of the van Hove function was calculated also from all sampled trajectories for all mobile ions together. To compute a "survival probability" or a "relaxation time" (Figure 2e and equivalents), the computed self part of the van Hove function was integrated radially to a threshold value, 0 to 1.7 Å for a single hop, and 0 to 4.6 Å for a double hop.

The displacements of individual ions were then averaged to yield each ion's time-averaged mean-square displacement (tMSD). Subsequent ensemble-averaging all single-ion tMSD in two dimensions within the conduction plane yields the plots in Figure 2ab and equivalents (Extended Data Figures 1-6). Simultaneously, the fourth power of the displacement was calculated and time-averaged (bar notation). The non-Gaussian parameter (NGP) was calculated for in-plane dimensionality *d*=2 as:

$$NGP(t) = \frac{d}{d+2} \frac{\overline{\langle r^4(t) \rangle}}{\overline{\langle r^2(t) \rangle}^2} - 1$$

Because transport is effectively two-dimensional at timescales corresponding to ≥ 1 hop, the non-Gaussian parameter approaches an asymptotic non-zero value at long time lags, but does not approach zero, if the out-of-plane dimension is included.

The diffusion kernel correlation $C_D$ was calculated following Song et al.[16] using the Laplace transforms of the ensemble-averaged tMSD (Figure 2a and equivalents) and of the fourth power of the time-averaged displacements. First, the Burnett correlation function[16–19] was computed as the second derivative of the fourth cumulant of displacement with respect to the time lag:

$$BCF(t) = \frac{1}{4!\,d}\frac{d^2}{dt^2}\left(\langle r^4(t)\rangle - 3\langle r^2(t)\rangle^2\right)$$

With $z$ as the Laplace variable and dimensionality $d=2$, the Laplace transform $\mathcal{L}$ of the diffusion kernel correlation $C_D$ in the Laplace space is:

$$\mathcal{L}\{C_D(t)\} = \frac{1}{2\,z^2\mathcal{L}\{\langle r^2(t)\rangle\}^2}\left[\frac{24\,\mathcal{L}\{BCF(t)\}\,d^2}{(d+2)\,z^2} + \mathcal{L}\left\{\langle r^2(t)\rangle^2\right\} - 2z\,\mathcal{L}\{\langle r^2(t)\rangle\}^2\right]$$

The diffusion kernel correlation is transformed back into the time domain using the Stehfest inversion algorithm[20]. Due to the noise in the simulated $\langle r^2\rangle$ and $\langle r^4\rangle$ (Figure 2b, Figure S1, and Extended Data Figures 1-6), and to the fact that $C_D$ takes near-zero values before crossing the origin, we do not assign a single point to define an exact time for $C_D \to 0$. Instead, Figure 5 shows conservative ranges.

The ergodicity breaking parameter EB[21] was calculated for a small time lag $t = 20$ ps and varying subsets of the overall simulation length (starting at the beginning of the simulation) as:

$$EB(t,\Delta) = \frac{\langle r^4(t,\Delta)\rangle - \langle r^2(t,\Delta)\rangle^2}{\langle r^2(t,\Delta)\rangle^2}$$

Song et al. proposed[16] that the relaxation of dynamic heterogeneity $C_D \to 0$ necessarily precedes $EB \propto \Delta^{-1}$. This holds for all simulated materials, and most clearly in Ag β-alumina (Extended Data Figure 4), where at 600 K the former occurs at $t \approx 10$ ns, and the latter at $\Delta \approx 20$ ns. We propose that tests for $C_D \to 0$ and $EB \propto \Delta^{-1}$ can serve as benchmarks for the extraction of constant-valued transport coefficients and DC-limit conductivities from large-scale simulations of ion conductors.

Center-of-Mass Diffusion Coefficient and Conductivity

The center of mass displacement of the mobile ions is recorded directly in LAMMPS. It is referenced to the rest frame of the host lattice by subtracting the lattice center-of-mass displacement[22]. Here, we denote the corrected displacement of the center-of-mass of mobile ions as $r_{CoM}$. For dimensionality $d=2$, and time lag $t$:

$$D_{CoM}(t) = \frac{\overline{r_{CoM}^2(t)}}{2td}$$

Here, the over-bar denotes time-averaging over the unique trajectory of the center of mass. The sampling of trajectories for each time lag is done with multiple starts 10 ps apart. While the trajectories sampled this way are not strictly independent, the results are not different for sampling strictly independent trajectories[15] where each of multiple starts is at least $t$ apart. The distributions of center-of-mass displacements sampled this way are exponential (Extended Data Figure 8) even at high temperature when ergodicity is reached and a DC limit could be expected. The longest time lag used is 2.5 ns, i.e. 1/40 of the duration of the simulation. The (in-plane) conductivity $\sigma(t)$ for a time lag $t$ is calculated via the Einstein relation using the elementary charge $q$, temperature T, and $N$ mobile ions within the simulation volume $V$:

$$\sigma(t) = \frac{1}{V}\frac{(qN)^2}{k_B T} D_{CoM}(t)$$

This is plotted in Figure 3, but sampled with the independence of sampled trajectories enforced, versus inverse time lag $t$, which corresponds to a frequency. Given that $r_{CoM}^2(t)$ is distributed exponentially for all time lags used (Extended Data Figure 8), the expectation value of $D_{CoM}(t)$ also characterizes the standard error. If the distribution were to have a different shape, the standard errors would need to be noted individually for each time lag.

### Hopping Statistics

Coordinates within the simulated conduction planes are assigned to Beevers-Ross and anti-Beevers-Ross sites based on a 2D Voronoi tessellation using O(5) oxygens as vertices of the Voronoi polygons. For the trajectory of a mobile ion, a hopping event is recorded every time the site whose center is closest to the ion changes, using a bounded-box query algorithm implemented in the *freud* python module[1]. This is a simpler hop detection algorithm than others[7,23], and in principle it could overcount "incomplete" hops with sub-picosecond residence times. Since we focus on the behavior at longer timescales, "incomplete" hops do not affect any results or conclusions. This approach also simplifies the recording of back-hops for the analysis of dynamic relaxation (Figure 5 and Extended Data Figure 9). Within a series of hopping events by an ion, diffusion events are hops that are followed by another hop to a new (third) site, while hops that are immediately reversed do not directly contribute to diffusion[24].

### Hopping Locations

As with the placement of defects, each conduction-plane mobile-ion site (i.e. the Voronoi polygon) is referenced by its distance to the nearest defect (for β) or by the number of defects adjacent to it above and below the conduction plane (for β``). Each hop is mapped to an origin site and a destination site. The non-returning hops, i.e. hops from site A to site B followed by a hop to site C, are diffusion events. The last hop for each ion within a simulation was discarded, as its outcome is indeterminate. The 2D matrices in Figure 4b-d and Figure S6b-d show the rates of diffusion events by their origin (horizontal axis) and destination (vertical axis) for β``-aluminas at 600 K.

### Site Occupancy

The site occupancies are calculated for every site by summing the residence times of all the hops into that site, plus the initial occupancy starting at time zero and ending with the first hop out of the site, and dividing by the total time of the simulation. For simulation durations where ergodicity is verified using $C_D \to 0$ and $EB \propto \Delta^{-1}$ (600 K for β``-aluminas, and 1000 K for β-aluminas), the occupancies can be assumed to have reached a thermal steady state. In that case, the occupancies of each site reflect a Maxwell-Boltzmann distribution of Helmholtz free energies, with each site as a distinct state with degeneracy 1. This includes both ion-host and ion-ion interactions. This quantity is plotted in Figure 4e-g, dis-aggregated by the "geographic" site descriptor (the count of neighboring defects) for β``-aluminas at 600 K. For β-aluminas, the corresponding calculation is performed at 1000 K (Figure S6) given that the distributions of trapping times near defects extend throughout the simulation at lower temperatures (Figure S7) and ergodicity cannot be assumed to be reached at 600 K and 300 K. If the calculation is attempted when ergodicity does not hold (simulation length too short for the temperature), the difference in energies between e.g. a-BR and BR sites is underestimated.

# Supplementary Information

Supplementary Note 1: Long-Time Behavior of the Non-Gaussian Parameter

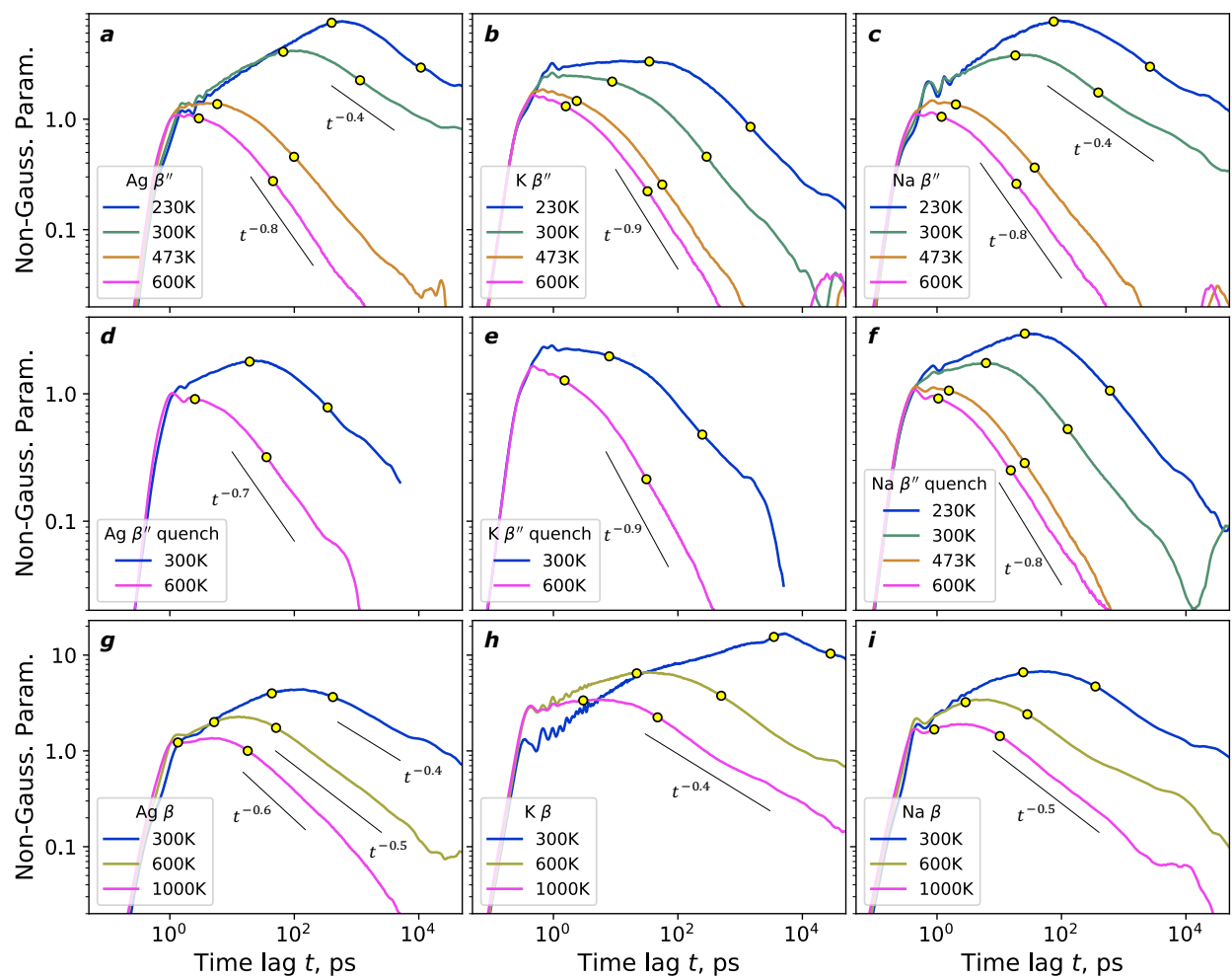

**Figure S1 | Long-Time Behavior of the Non-Gaussian Parameter. a-c**, β''-aluminas. **d-f**, β''-aluminas with a distribution of defects simulating quenching[8,14]. **g-i**, β-aluminas. **a,d,g**, Ag. **b,e,h**, K. **c,f,i**, Na. The long-time regime is characterized by a power-law decay with approximate guides noted. For β''-aluminas in the low-temperature regime and for β-aluminas, the exponent ≈-0.5. For β''-aluminas in the high-temperature regime, the exponent increases towards ≈-1.

Since at short timescales the distributions of displacements are wide-tailed, the non-Gaussian parameter (NGP) should yield information on the evolution of dynamic heterogeneity[25,26]. In all β''-aluminas (Figure S1a-c), the NGP peaks close to the 1-hop relaxation time (shorter in the high-temperature regime), and reaches a long-time regime with a power-law decay starting between the one-hop and two-hop relaxation times. For Na and Ag, the exponent in the long-time regime switches from ≈-0.4 at ≤300 K to ≈-0.8 at ≥473 K. The doubling of the long-time exponent coincides with the switch between the low- and high-temperature activation regimes in conductivity. This differs from K β''-alumina (Figure S1b), or β''-aluminas with a distribution of defects that mimics quenching (Figure S1d-f)[8,14], where the NGP has long-time exponents ≈-0.9 and ≈-0.8, respectively, at all simulated temperatures. In β-aluminas (Figure S1g-i), the NGP peaks between the 1-hop and 2-hop relaxation times, and the long-time exponent is ≈-0.5 at all temperatures.

These simulation results are consistent with at least one model of sub-diffusive dynamics in that the convergence of the NGP to zero proceeds with an exponent less than unity[27]. However, the exponent of the NGP relaxation does not match the exponent of the tMSD, so such a model is not fully consistent with the dynamics of β- and β``-aluminas. Song et al.[16] predict the NGP to decay as $t^{-1}$ at long times, unless the waiting-time distribution (here, hop residence time distribution) is long-tailed. The distribution of hop residence times is indeed long-tailed (Supplementary Note 3), and mobile-ion vacancy ordering (β``-aluminas, Supplementary Note 2) or clustering at defects (β-aluminas, Supplementary Note 4) add to such heterogeneity. Overall, the long-time regime NGP suggests long-lived heterogeneity, and is consistent with other simulation results. However, the NGP shows fewer features and is less interpretable than $C_D$. Notably, unlike the NGP, $C_D$ can exhibit multiple peaks, highlighting multiple timescales or mechanisms of heterogeneity (compare e.g. Extended Data Figure 2f versus Figure S1b for K β``-alumina). A more detailed comparison is left for the future.

Supplementary Note 2: The Origin of Mobile-Ion Vacancy Ordering in β``-aluminas

The low-temperature regime of activation in β``-aluminas is distinguished by the ordering of mobile-ion vacancies, detectable via diffuse scattering. The diffuse scattering signature disappears on heating, concurrent with the decrease in activation energy[28–31]. Our simulation reproduces the ordering of mobile-ion vacancies in the expected superlattice with unit cell dimension $\sqrt{3} \times \sqrt{3}$ (Figure S2ab), and its basis vectors rotated by 30 degrees from the crystallographic axes. By inspection of Figure S2, the mobile-ion vacancies within the superlattice exclusively occupy conduction-plane sites without "direct" $Mg_{Al}$ defect neighbors (signified by green dots in the middle of the sites), although not always the same exact sites (compare 230 K and 300 K simulations, Figure S2ab). The sites with "direct" $Mg_{Al}$ defect neighbors evidently disrupt the ordering of mobile-ion vacancies. These sites have consistently lower average Helmholtz free energies, calculated at 600 K when ergodicity holds, i.e. they attract mobile ions and repel vacancies.

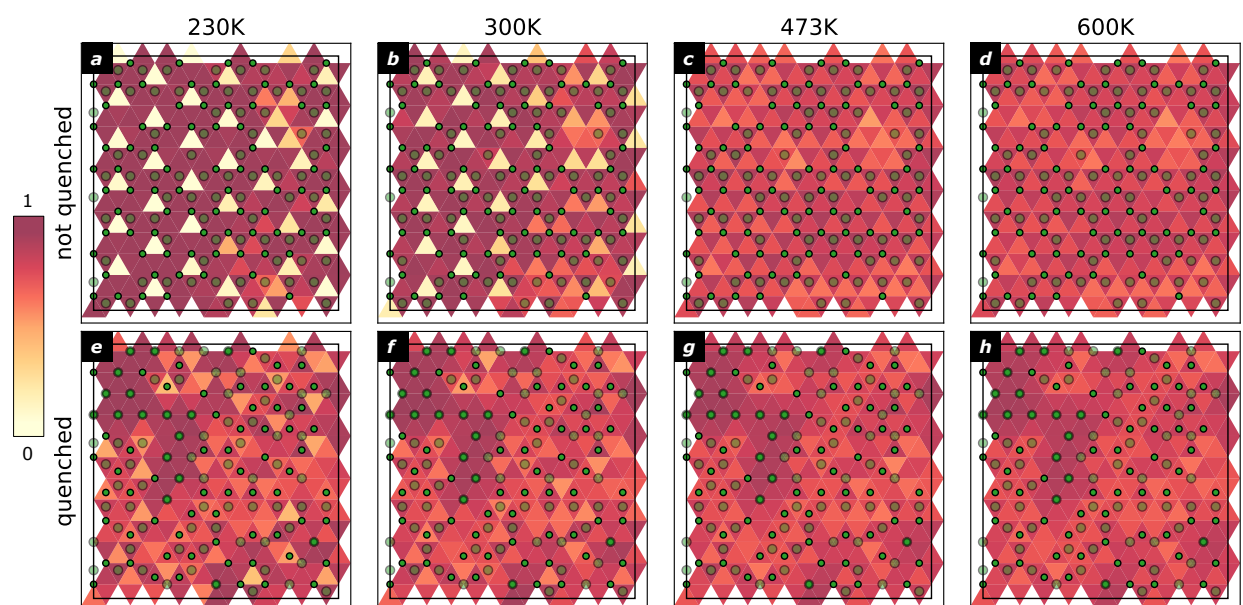

**Figure S2 | Sites without "direct" $Mg_{Al}$ defect neighbors anchor the ordering of mobile-ion vacancies in Na β"-alumina.** Simulated occupancies of mobile-ion sites in a representative conduction plane for temperatures (a) 230 K, (b) 300 K, (c) 473 K, (d) 600 K. Each conduction-plane site is shown as a triangle. Green dots: $Mg_{Al}$ defects above (smaller, opaque green dots) and below the conduction plane (larger, dimmer green dots). "Direct" $Mg_{Al}$ neighbors are dots in the centers of the conduction-plane sites, which sit on one side of the plane in (**a-d**), but on both sides in (**e-h**). The superlattice of mobile-ion vacancies (low-occupancy sites, exclusively without "direct" $Mg_{Al}$ neighbors) is obvious at 230 K and 300 K. Corresponding occupancies for a distribution of $Mg_{Al}$ reflecting quenching: (e) 230 K, (f) 300 K, (g) 473 K, (h) 600 K. The superlattice is absent.

A supercell of the mobile-ion vacancy superlattice contains six sites, therefore there are six sets of sites for possible superlattices. In the simulated "un-symmetric" distribution of $Mg_{Al}$ defects (Figure S2a-d), representative of slowly-cooled crystalline material[8,14], "direct" $Mg_{Al}$ defect neighbors sit on one side (above / below) the conduction plane. All conduction-plane sites with such "direct" neighbors belong to only three of the six possible sets of sites for the superlattice. This can be seen in Figure S2a-d: all sites with "direct" $Mg_{Al}$ defect neighbors are downward-pointing triangles, separated from each other by multiples of two hops. We conjecture that the reason for the repulsion of vacancies from sites with "direct" neighboring defects stems from the "un-symmetric" distribution of defects in non-quenched material. These conduction-plane sites have nearest-neighbor defects on both sides of the plane (Figure 4a, Figure S2a-d), canceling at least in part the out-of-plane Coulombic forces. Sites with only "offset" defect neighbors only have nearest defects on one side of the plane, which could leave the Coulomb interaction un-mitigated. In quenched material (Figure S2e-h), this distinction is scrambled, and the Coulomb interaction becomes more balanced on average.

In non-quenched material, the vacancy superlattice remains free to form without disruption on three conduction-plane sub-lattices without "direct" defect neighbors (upward-pointing triangles in Figure S2a-d). Similarly, these are also separated by two hops from each other. This implies that a boundary between two vacancy superlattices can shift via double hops. This is most easily seen in the right side of the lattice in Figure S2b, which captures a "grain boundary" between two ordered regions (omitting the periodic boundary conditions of the simulation). A double hop at the edge of two vacancy superlattices will essentially move one vacancy from one to the other. This double-hop vacancy diffusion at the boundary of mobile-ion superlattices must be the origin of the low-temperature two-hop diffusion mechanism in β``-aluminas deduced from the time dependence of $C_D$ (Figure 2f). This carries three implications.

First, the Helmholtz free energy difference between "direct" and "offset" conduction-plane sites, ≈40 meV in Na β``-alumina, controls the temperature stability of the ordering of mobile-ion vacancies. Once this energy difference is comparable to thermal energy, the ordering can no longer be stable. This is consistent with the average Helmholtz energy difference in Figure 4 matching $k_BT$ for the temperatures (≈500 K) at which the superlattice disappears, and the activation energy switches between low-temperature and high-temperature regimes. The gradual nature of the switch is also consistent with the disorder in the Helmholtz free energies seen in Figure 4.

Second, at low temperatures the stability of the superlattice sets the timescale for the breaking of ergodicity and the non-equivalence of mobile-ion trajectories. At 230 K, a few Na ions do not attempt even one hop in the first 100 ns, necessitating a longer simulation (300 ns). The trajectories become interchangeable, and ergodicity is reached, only once every ion has sampled being both a part of the superlattice, and a part of its mobile boundary. The slow exchange of ions between the superlattice and the boundary is the origin of ergodicity breaking, and the origin of the frequency dispersion in conductivity at frequencies slower than the Jonscher regime.

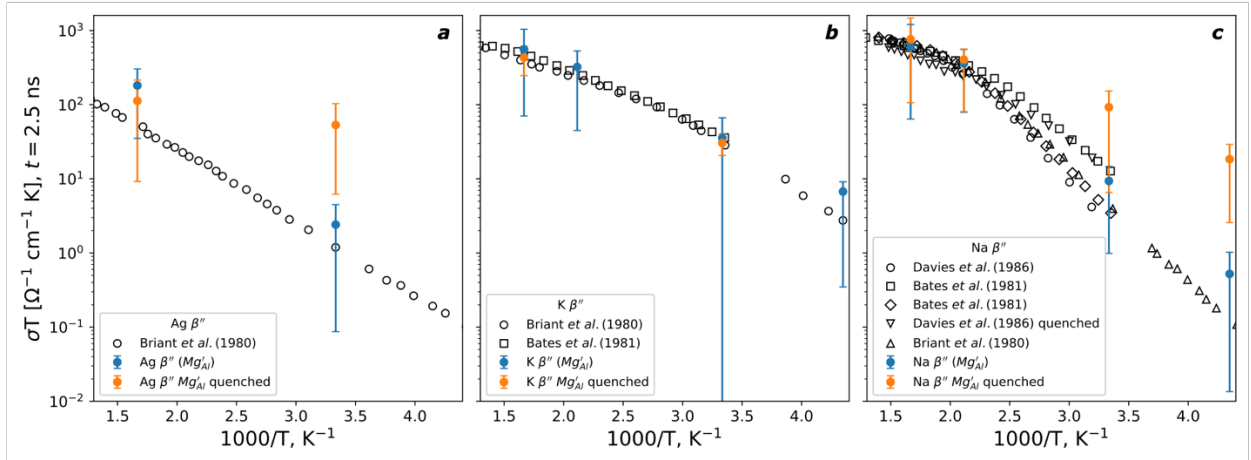

**Figure S3 | Arrhenius relations for β''-aluminas as a function of process-modulated defect distributions. a**, Ag. **b**, K. **c**, Na. Simulations with defects distributed un-symmetrically (blue) as shown by Davies *et al.*[14] using diffraction, and randomly as shown to arise from quenching (orange). Black-contour symbols: experimental conductivity values[14,32,33]. The error bars represent the range between $10^{th}$-$90^{th}$ percentiles of the distributions of eMSD. Simulation length $\Delta = 100$ ns except for the quenched Ag and K simulations, where $\Delta = 10$ ns.

Third, a quenching process that yields a "symmetric" distribution of defects[8,14] spreads the "direct" $Mg_{Al}$ defect neighbors across all six sets of possible locations for the superlattice. This equalizes the energetics of the superlattice forming on any one of them, and disrupts the formation of the mobile-ion vacancy superlattice, at least for any extended period of time or region of space. This is responsible for the increase in conductivity upon quenching observed experimentally in Na β``-alumina, and reproduced in our simulations (Figure S3).

## Supplementary Note 3: Wide-tailed Distributions of Hop Residence Times

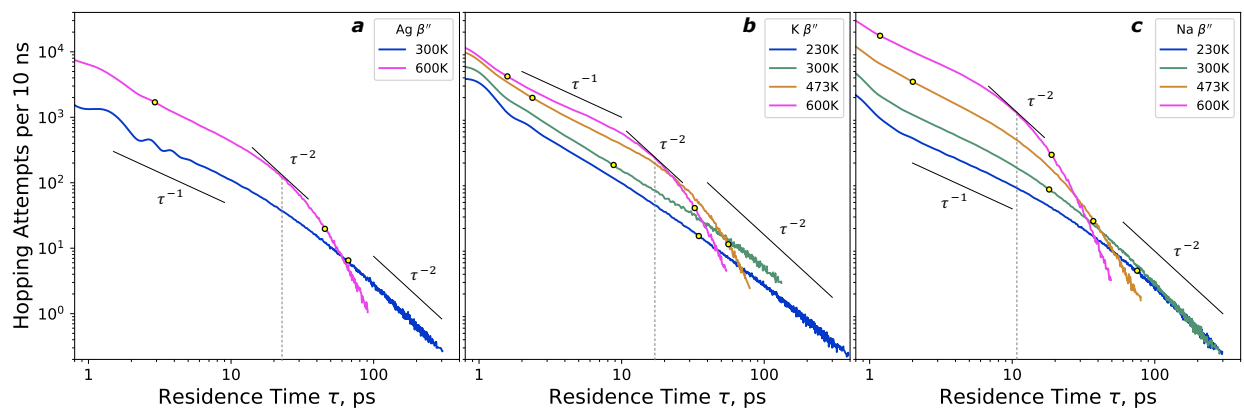

**Figure S4 | Wide-tailed distributions of residence times (τ) between hops in β''-aluminas. a**, Ag, **b**, K, **c**, Na. Yellow circles represent one-hop and two-hop relaxation times from the self part of the van Hove function. At high temperatures, the timescale for $C_D \rightarrow 0$ approximates the shortest residence times at which the distributions $\propto \tau^{-2}$, and an expected residence time $\langle \tau \rangle$ can be defined (shown for 600K). Hopping attempts are events recorded when an ion crosses a crystallographic site boundary. The distributions plotted here are smoothed with a Gaussian filter of 200 fs HWHM, truncating the dsitribution. The longest residence times for singular rare-event long-lived hops are highlighted using log-scale binning by τ instead in Figure 5.

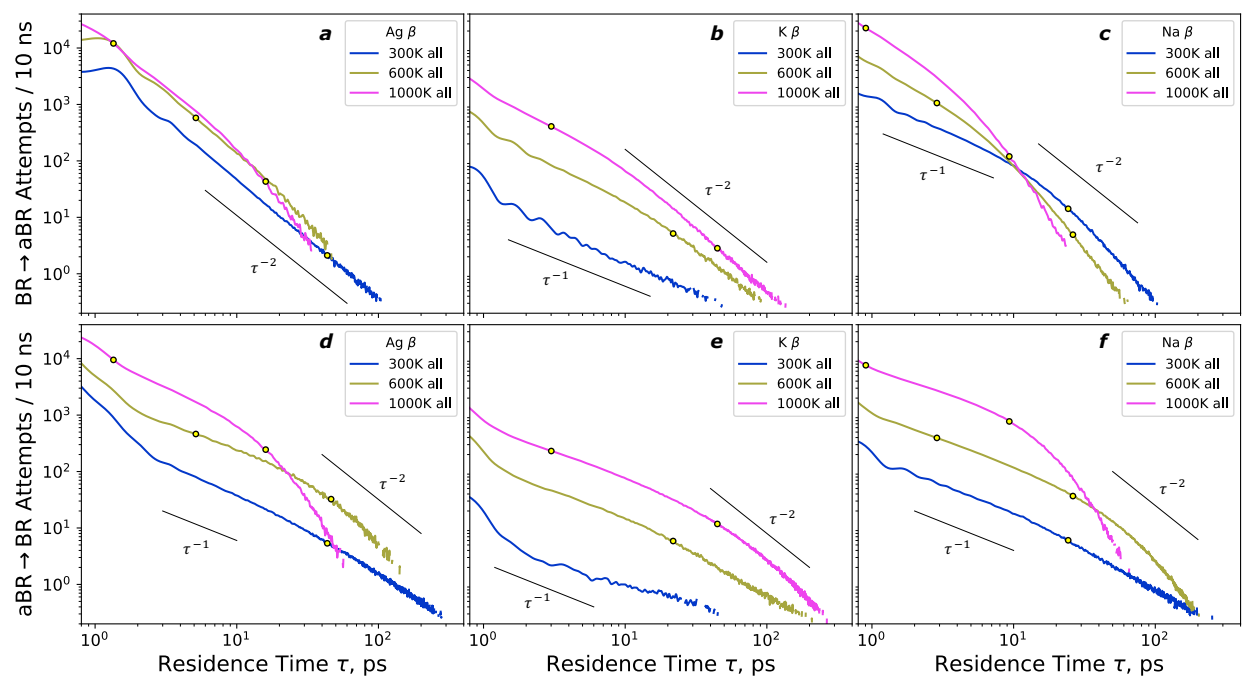

**Figure S5 | Distributions of residence times (τ) between hops in β-aluminas. a-c**, from BR sites to aBR sites, **d-f**, from aBR sites to BR sites. **a** and **d**, Ag, **b** and **e**, K, **c** and **f**, Na. Yellow circles represent one-hop and two-hop relaxation times from the self part of the van Hove function. Hopping attempts are events recorded when an ion crosses a crystallographic site boundary. The distributions plotted here are smoothed with a Gaussian filter of 200 fs HWHM.

The quantity most analogous to a waiting-time distribution of a continuous time random walk in our simulation is the distribution of hopping residence times $\tau$. Here, we record all site-boundary crossings by mobile ions irrespective of their outcome (back-hop or diffusion event), and treat the correlation factor separately (Figure 5). The distributions of residence times $\tau$ between such site-boundary crossings, i.e. the waiting-time distributions $\psi$, have a power-law distribution for short times: $\psi(\tau) \propto \tau^{-1}$ (Figures S4 and S5). This signifies the presence of a waiting-time memory, as for a continuous time random walk[34], in addition to the low values of the correlation factor. Furthermore, for short residence times, $\psi(\tau)$ is a wide-tailed distribution with an undefined first moment $\langle\tau\rangle$. For longer residence times (larger than two-hop relaxation times), $\psi(\tau)$ decays at least as $\tau^{-2}$, and at least the first moment $\langle\tau\rangle$ can be defined. For β``-aluminas (Figure S4), for the simulation length 100 ns, the second moment $\langle\tau^2\rangle$ is only definite in the high-temperature regime, and only for residence times larger than at least the timescale $C_D \to 0$. Like the correlation factor $f$ (Figure 5), the shape of $\psi(\tau)$ also suggests that the kinetics of hopping at room temperature are correlated. However, the waiting-time distribution and the correlation factor are distinct types of memory, and could be influenced by the same or distinct mechanisms.

Pinpointing the onset of and the atomistic mechanism behind the Jonscher regime of AC conductivity would be a breakthrough for predicting practical performance of materials. For Na β``-alumina, where AC conductivity measurements are available (Figure 3a), the residence times $\tau$ above which $\psi(\tau)$ decays at least as $\tau^{-2}$, and $\langle\tau\rangle$ can be defined, approximate the lowest frequency of the Jonscher regime. These are: ≈100-250 ps at 230 K, ≈40-50 ps at 300 K, ≈15-18 ps at 473 K, and ≈10-12 ps at 600 K. For Na and Ag β-aluminas (Figure 3bc and Extended Data Figures 1 and 4), the two-hop relaxation time provides the closest approximation of the timescales of the Jonscher regime. While we are unable to propose a universal descriptor for the extent of the Jonscher regime, we have shown that a sizeable fraction of mobile ions completes diffusion events even within the Jonscher regime, and some complete multiple diffusion events. Our results point to a non-ergodic, dynamically heterogeneous process of ionic conduction within the Jonscher regime, i.e. an ill-defined $\langle\tau\rangle$ and/or location- and a timescale-dependent correlation factor $f$ for hops and diffusion events (Figure 5). We conjecture that the Jonscher regime signifies a viscoelastic rather than necessarily localized mode of transport. Sub-diffusive transport within the Jonscher regime could be analogous to fractional Langevin dynamics with anticorrelated noise and a finite memory kernel[35,36] giving rise to the timescale dependences of the correlation factor $f$ (Figure 5) and diffusion kernel correlation $C_D$ (Figure 2f).

Supplementary Note 4: Site Energetics and Formation of Defect Clusters in β-aluminas

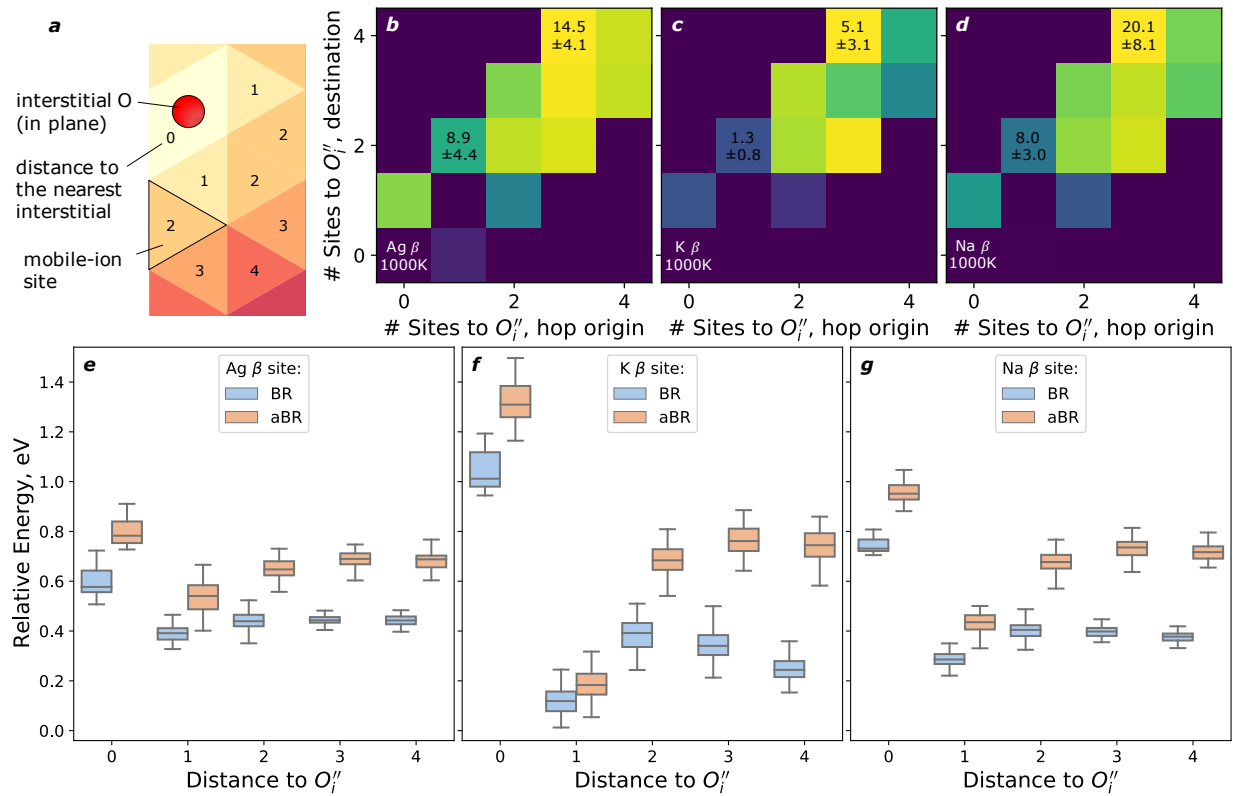

**Figure S6 | Site energetics in β-aluminas. a**, Schematic of the site distances (numbered) from the oxygen interstitial (red) within the conduction plane of β-aluminas. **b-d**, Rates of diffusion, in units of non-returning BR → aBR hops per edge and per nanosecond, classified by their sites of origin (horizontal axes) and destination (vertical axes), in β-aluminas at 1000 K. All color scales are normalized to the noted maximum rates (yellow). The uncertainty value is a standard error over the edges that fit this criterion. Also noted are the rates of de-trapping as the non-returning hops from sites at distance 1 to sites at distance 2. **e-g**, Distributions of relative Helmholtz free energies for all conduction-plane sites β-aluminas at 1000K, aggregated by distance from the nearest oxygen interstitial (horizontal axis), and for Beevers-Ross (BR, blue) vs anti-Beevers-Ross (aBR, orange) sites. The sites at distance 0 are those at which edges the interstitials reside. These sites are partially sterically blocked and have large energies. Both BR and aBR sites at distance 1 are nearly fully occupied. The Helmholtz free energy differences between non-defect-adjacent BR and aBR sites, e.g. median 2-aBR to median 3-BR, are 0.20 eV, 0.32 eV, and 0.28 eV for Ag, K, and Na, respectively. All median BR-aBR energy barriers are higher than the activation energies; the agreement is closest for Ag β-alumina.

Using the same method as for β''-aluminas, at high temperature (1000 K) when ergodicity can be verified with $C_D \to 0$ and $EB \propto \Delta^{-1}$ (Extended Data Figures 1, 4, 6, and 7), we use the time-averaged site occupancies for β-aluminas to calculate the location-dependent diffusion rates and Helmholtz free energies (Figure S6). The sites immediately sharing an oxygen interstitial (distance to $O_i$ = 0) are sterically blocked and have high energies, and the sites next to them have low energies because of Coulombic attraction (distance to $O_i$ = 1). These sites house the mobile ions that are part of the cluster with the oxygen interstitial. Away from the interstitials, the aBR sites have lower occupancies and higher energies than the BR sites. As expected, diffusion occurs at sites with distances to $O_i \geq 2$. Additionally, for K β-alumina with the largest mobile ion, the sites with distances $\geq 4$ appear to be less active for diffusion (Figure S6b, rightmost column), and diffusion can be thought to proceed via the "edge" of defect clusters (distances 2 and 3) where the full occupancy of the cluster sites (distance 1) could be perturbing the nearby aBR bottleneck.

Notably, the calculation of average relative site energies must be made only at a combination of temperature and simulation length such that ergodicity is verified. If this is not done, the energy difference is underestimated. The BR-aBR energy difference can be approximated using medians of all single-site energies in our simulation of the same type and location (for sites sharing edges, e.g. BR at distance = 2 versus aBR at distance = 3). However, for all three materials simulated, this estimate is higher than the activation energy for conductivity. The closest match between activation energy and Helmholtz free energy is for Ag β-alumina (Figure S6e), where our simulation also reproduces experimental AC and DC conductivities the closest (Figure 3). The mismatch may be due to a combination of (1) the error in the classical interatomic potentials in describing the two-coordinate aBR site, and (2) the contribution of local, non-equilibrium configurations to conduction. The latter is most commonly termed the "knock-on" effect and is conceptually similar to interstitialcy conduction. A stoichiometrically excess mobile ion necessarily occupies a high-energy site (here, aBR). Its proximity perturbs the energy and position of the nearest-neighbor mobile ions in the low-energy sites (here, BR) and enables the BR → aBR hops with a lower activation energy (and likely a distinct attempt frequency) than for a stoichiometric material. The interstitialcy-like "knock-on" mechanism of hopping in β-aluminas, where a diffusion event requires a double-hop BR → aBR → BR sequence, is also the origin for the peak of $C_D$ occurring between one-hop and two-hop relaxation times at all temperatures (Extended Data Figures 1, 4, and 6).

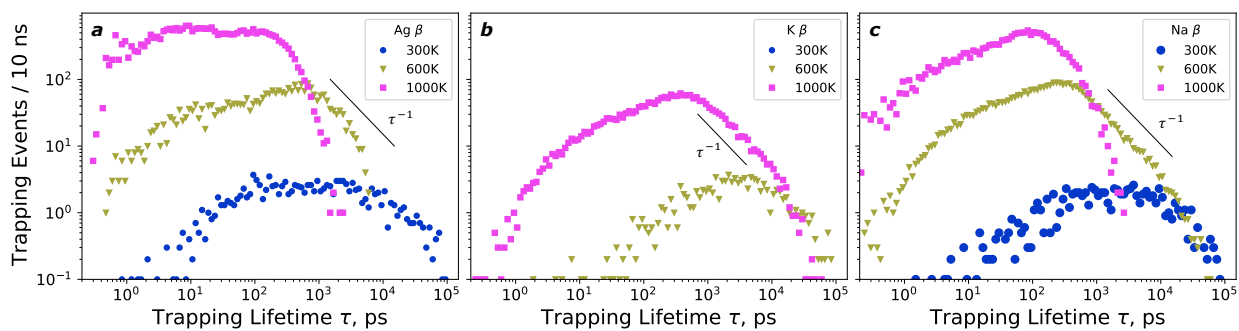

**Figure S7 | Kinetics of trapping of mobile ions by oxygen interstitials in β-aluminas. a**, Ag. **b**, K. **c**, Na. The distributions of all trapping events (vertical axis), when a mobile ion enters a cluster (trapping) and later leaves (detrapping), are aggregated into logarithmically spaced bins by the residence time of the mobile ion within the cluster (horizontal axis). All simulation lengths are 100 ns; the statistics are clearly limited by the simulation lengths at 300 K.

The formation of a cluster of mobile ions (M) with $O_i$ follows the chemical reaction:
$$O_i + 4M \rightarrow [M_4 O_i]$$
Sites at distance = 1 (both BR and aBR, Figure 1f) from the interstitial house the cluster of four mobile ions, which are bound and unable to diffuse for extended residence times. These sites are the most fully occupied, and have the lowest Helmholtz free energies (Figure S6) in the material. The exchange of metal ions between the bound and free states should bear on the simulation reaching ergodicity, and on the frequency-dependent AC conductivity. We quantify the kinetics of this exchange of metal ions between the cluster and all other, unbound or free, locations, using the simulation times elapsed between a mobile ion becoming trapped, and the same ion leaving the cluster. The trapping is a hop sequence starting at a site with distance = 2 or 3, via a site with distance = 2 to a third site with distance = 1. Once the mobile ion arrives at the site with distance = 1, it is a part of the cluster and is trapped. It may attempt to hop to sites with distance = 0 or distance = 2 within the cluster many times. However, de-trapping requires the reverse of the trapping sequence: hops from a site at distance = 1 to a site at distance = 2 and onward to a third site at distance = 2 or 3. The largest peaks in the distributions of trapping times match the average de-trapping rates in Figure S6b-d, for example ≈100-150 ps versus ≈8-9 $ns^{-1}$ for Na and Ag. The longest residence times of metal ions within the defect-bound clusters are up to ≈1 ns at 1000 K for Na and Ag. This timescale matches the timescales for $C_D \to 0$ for Na and Ag β-aluminas at 1000 K (Extended Data Figures 1 and 4). This verifies that the slow exchange between bound and free states is responsible for the long-time regime of $C_D$ and for the dispersion in mobile-ion transport at timescales longer than the two-hop relaxation time. Furthermore, ergodicity is only reached at timescales longer than this exchange, once every ion within the simulation has sampled its two possible states: bound in a cluster, and free. A simulation shorter than the timescale of this exchange would effectively yield two split populations of mobile ions. This is similar to single-particle tracking experiments, which may not offer trajectories of a sufficient length to observe an exchange between heterogeneous populations, and thus observe population splitting[36] instead.

At temperatures lower than 1000 K, the exchange becomes slower, and the sampled trapping and de-trapping events become more rare. The "lifetime" of the bound state becomes longer than the simulation length. Additionally, the tail of the distribution of residence times in the cluster becomes wider at lower temperature (flatter than the $\tau^{-1}$ power-law guidelines in Figure S7). This suggests that the first moment of the overall distribution of lifetimes, and the expected lifetime of the bound state, may be undefined at 300 K.

We are now able to ascertain the origin of sub-diffusion in β-aluminas. The exponent of sub-diffusive tMSD is temperature-dependent, and the sub-diffusive exponent persists for larger displacements than distances between defect clusters (Extended Data Figures 1,4,6), which are ≈10-20 Å. This makes a tortuosity mechanism[37], where ions navigate a longer path around defects than the total resulting displacement, insufficient to fully explain the simulated sub-diffusion. Given additionally that (1) the two-hop relaxation time matches the longest timescale of the Jonscher regime at 300 K (300-500 ps or 2-3 GHz in Na and Ag β-aluminas), (2) the formation of clusters breaks ergodicity at timescales longer than the two-hop relaxation timescale, and (3) the distribution of cluster lifetimes may be unbounded at 300 K, we conclude that the formation of long-lived clusters is the origin of the macroscopic long-time, low-frequency dispersion in AC conductivity, $\sigma \propto v^{0.1}$ (Figure 3). The values of tMSD at which the transport becomes Fickian in our simulation are e.g. ≈40-50 Å at 1000 K for Na β-alumina (Extended Data Figure 1), and are expected to be larger at lower temperatures. Within a model that does not incorporate arbitrarily long trapping times, such as by Kamishima *et al.*[38], these tMSD values correspond to a correlation length. Our simulation and our interpretation of it are thus consistent with their experimental result.

Supplementary Note 5: Geometric Crowding by Oxygen Interstitial Clusters

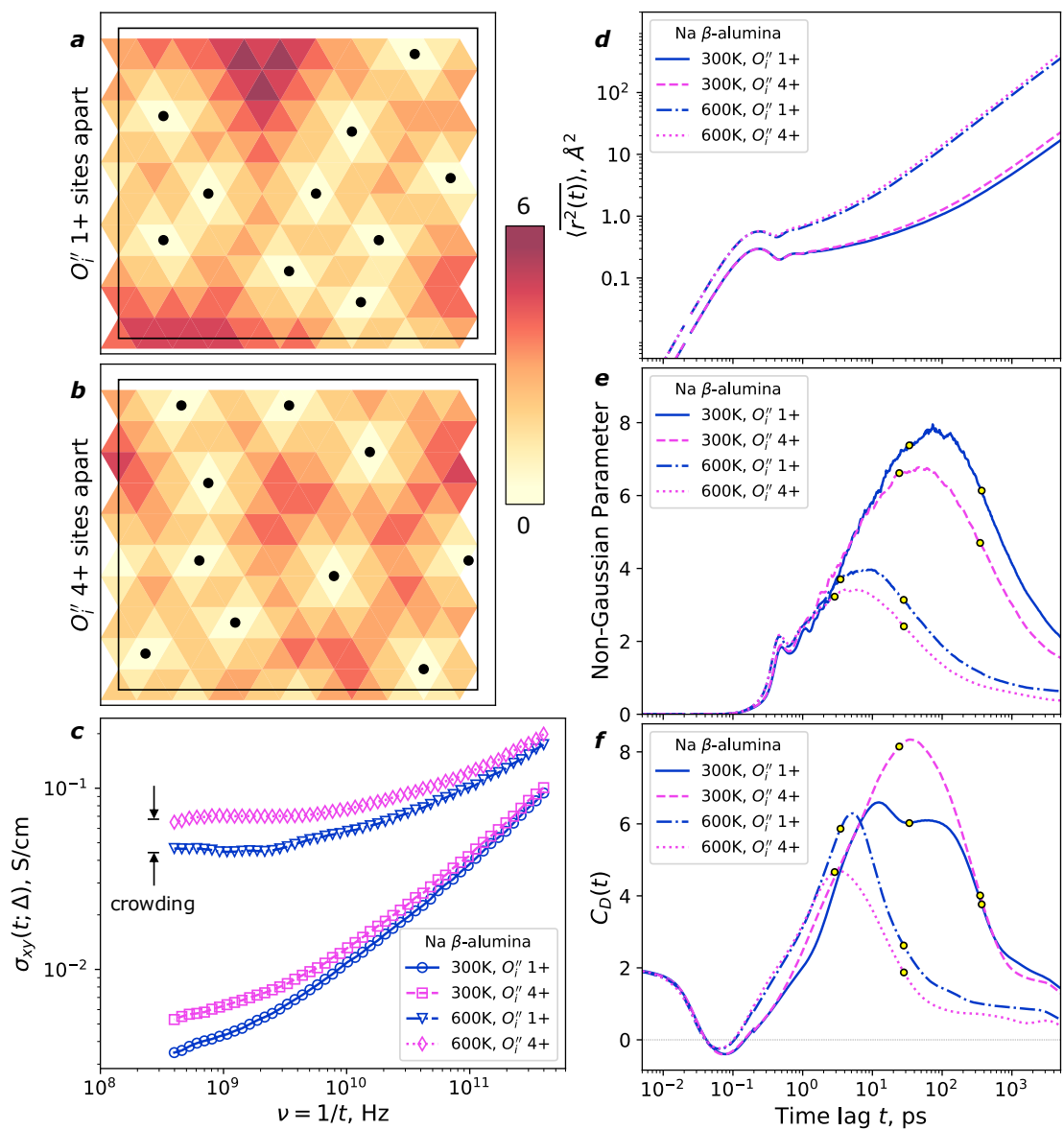

**Figure S8 | Effect of the distribution of oxygen interstitials on conduction in Na β-alumina.**
**a-b**, Maps of site distances to the oxygen interstitials (black dots) in one representative conduction plane for a simulation with the defects at least one site apart (**a**) or at least four sites apart (**b**). Sites at distance = 0 share the oxygen interstitials. In (**a**) and (**b**), each site is shown as a triangle; darker shading is farther away from a defect. **c**, Conductivity spectra for the two distributions of defects at 300 K and 600 K. The simulations with spaced-apart defects show consistently higher conductivities at lower frequencies. For the four simulations in (**c**): **d**, tMSD. **e**, Non-Gaussian parameter. **f**, Diffusion kernel correlation $C_D$. In (**f**), one- and two-hop relaxation times are shown as yellow circles.

In β-aluminas, the interstitial oxygens and the metal ions in clusters around them remain stationary on timescales comparable to multiple hops of otherwise mobile ions (Supplementary Note 4). The mobile ions must hop around the clusters, not through them. The clusters occupy 30% of the conduction-plane sites in our simulation. At this concentration of stationary obstacles, frequent reflections from the obstacles can locally slow down diffusion. As mobile ions diffuse, for example through a narrow pathway between nearly adjacent defect clusters, the correlation factor can be locally reduced by the fully occupied sites belonging to the clusters. This effect is termed crowding. For comparable concentrations of obstacles (30%), geometric crowding has been shown to affect diffusion in two-dimensional biophysical environments[39,40]. In particular, as crowding slows down a fraction of diffusing particles, it increases dispersion in the tMSD, and results in anomalous, non-Gaussian statistics. This effect is expected to arise when tMSD is sufficiently large to be comparable to the sizes of obstacles and distances between them. If such an effect were present in β-aluminas, it would be expected to affect conductivity at timescales when tMSD is at least as large as the size of a defect, i.e. at least longer than the two-hop relaxation times.

To verify the effect of the distribution of interstitial oxygens on transport, we simulate two quasi-random distributions of interstitials in Na β-alumina (Methods): located ≥1 site apart (more disordered), and located ≥4 sites apart (more ordered – and used for all simulations in the main text). While we are not aware of experimental evidence for any particular distribution of interstitials, the critical influence of processing-modulated $Mg_{Al}$ locations on the conductivity in β``-aluminas (Figures S2 and S3) motivates a similar inquiry. We conjecture that given the double negative charges on the interstitials, they should repel each other. A kinetically controlled distribution of interstitials would be disordered, with defects possibly locating close to one another and clumping. In our simulation, this is the distribution where interstitials are at least one site apart (Figure S8a). By contrast, thermodynamic control of material synthesis should yield a distribution where interstitials are well-separated. In our simulation, four sites is the largest distance possible to impose at the chosen $M_{1.2}$ stoichiometry without ordering (Figure S8b).

Figure S8c shows that simulated conductivities of Na β-alumina with interstitials 4+ sites apart ("thermodynamic control") are larger than that of Na β-alumina with interstitials 1+ sites apart ("kinetic control"). The magnitude of the difference is ≈2x at 300 K and 600 K. The effect is pronounced only at timescales ≥1 ns, larger than two-hop relaxation times. Within transition state theory, this effect is considered part of the Arrhenius prefactor, and not the activation energy. By contrast, the tMSD is nearly unaffected by the distribution of defects (Figure S8d): the effect of disorder pertains to eMSD more strongly. The non-Gaussian parameter (dispersion of mobile-ion tMSD) is larger for the disordered-interstitial simulations (Figure S8e). Notably, the diffusion kernel correlation at 300 K (Figure S8f) shows additional structure for the simulation with disordered defects at the timescale of one-hop relaxation. This is consistent with the existence of multiple environments with slightly differing kinetics of hopping. No other statistical descriptor enables such a distinction to be made.

All these features are consistent with geometric crowding. We conclude that a disordered distribution of interstitial oxygen defects, such as could arise from kinetically controlled, e.g. quenching, synthesis, is expected to be detrimental to conductivity in Na β-alumina. There is not yet experimental verification of this computational result. However, the difference in AC conductivities simulated in Figure S8c, in particular its frequency dependence, is similar to that between melt-grown and flux-grown crystals of similar total stoichiometry (Figure 3b)[41], and provides one possible explanation for the strong dependence of conductivity in β-aluminas on synthesis conditions. Of course, other defect reactions are also possible.

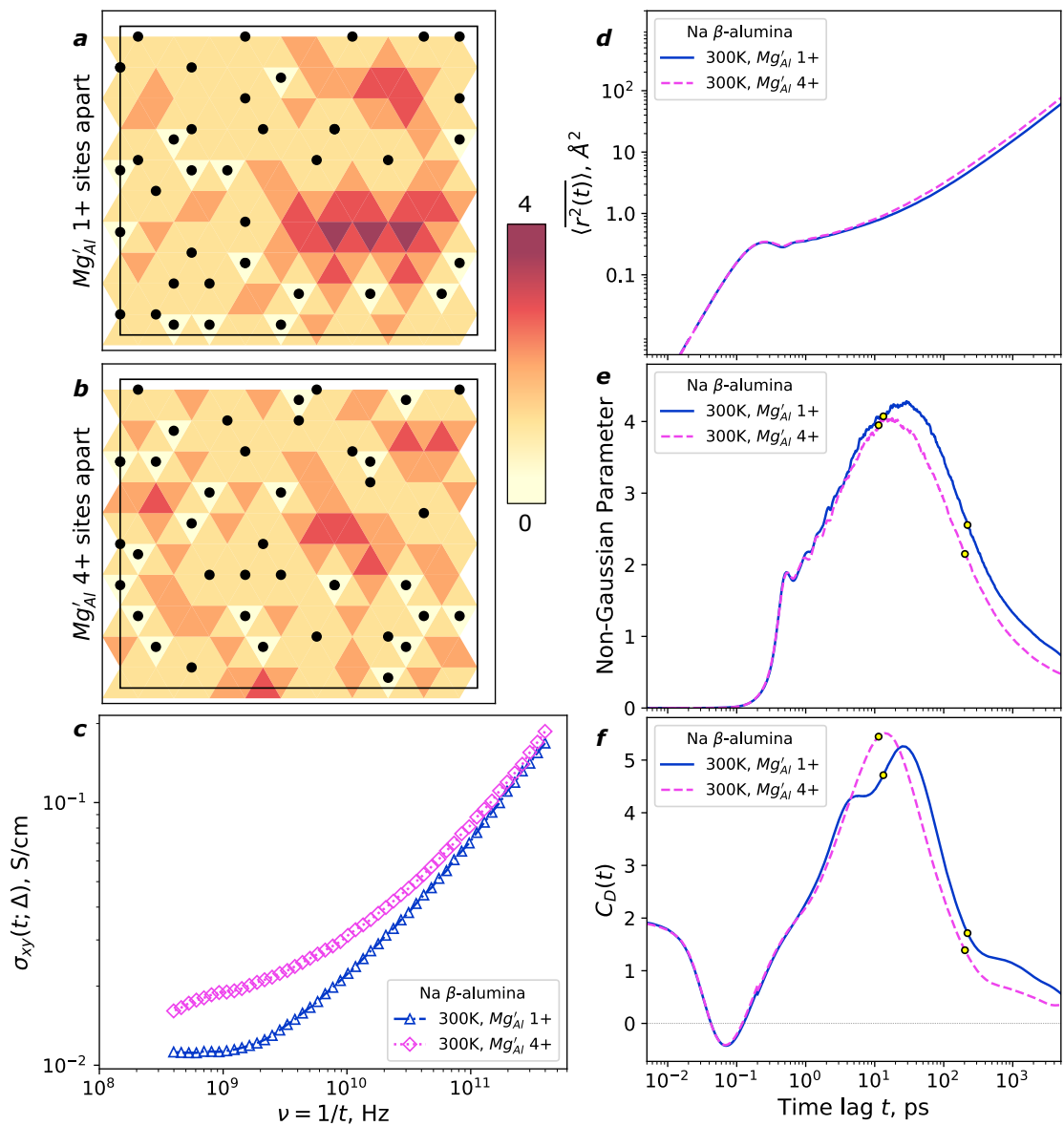

**Figure S9 | Effect of the distribution of Mg$_{Al}$ defects on conduction in Mg-doped Na β-alumina. a-b**, Maps of site distances to the Mg$_{Al}$ defects (black dots) in one representative conduction plane for a simulation with the defects at least one Al(2) apart (**a**) or at least four Al(2)'s apart (**b**). As in β``-alumina, defects both above and below the conduction plane are shown. In (**a**) and (**b**), each site is shown as a triangle; darker shading is farther away from a defect. **c**, Conductivity spectra for the two distributions of defects at 300 K. The simulations with spaced-apart defects show consistently higher conductivities at lower frequencies. For the two simulations in (**c**): **d**, tMSD. **e**, Non-Gaussian parameter. **f**, Diffusion kernel correlation $C_D$. In (**f**), one- and two-hop relaxation times are shown as yellow circles.

To verify the generality of the effect of crowding on conductivity, we simulate Mg-doped Na β-alumina[42–44] with differing distributions of $Mg_{Al}$ defects (Figure S9) for the same stoichiometry of mobile ions, $M_{1.2}$. As the network of Al(2) sites on which the Mg is expected to substitute mimics that of the mobile-ion sites, the two quasi-random distributions are 1+ Al(2) site apart (Figure S9a), and 4+ sites apart (Figure S9b). As for β``-aluminas, the defects in two neighboring spinel blocks affect the conduction plane between the blocks. If each $Mg_{Al}$ binds one mobile ion, then the concentration of sites inaccessible to diffusion becomes 40%. As with the interstitial oxygens, the more disordered distribution of defects yields a lower simulated conductivity (Figure S9c). All statistical descriptors (Figure S9d-f) show the same differences as with oxygen interstitials: a disordered distribution of defects leads to larger dispersion in tMSD, and the appearance of additional structure in $C_D$ on the timescales of hopping. We conclude that crowding also affects Mg-doped β-alumina via the stationary metal ions bound to defects. However, the overall simulated ionic conductivity is larger for Mg-substituted Na β-alumina, reflecting the weaker binding by each individual defect.

Supplementary Note 6: Trends with Mobile-Ion Size and Concentration

Here we highlight simple trends in conductivity with the identity of the mobile ions and dopants. The balance of repulsions between mobile ions and attractions between mobile ions and static defects determines the overall chemical trends.

The first factor of the balance is the repulsion between mobile ions, modulated by the size of the ions themselves. The potassium ion is the largest mobile ion. At sufficiently high concentration in the conduction planes, such that any mobile ion possesses repelling neighbors, $K^+$-$K^+$ repulsion dominates over $K^+$-host and $K^+$-defect interactions. This repulsion must be the driver of fast, site-independent diffusivity in K β``-alumina. In K β-alumina, the large size of K impedes conduction relative to the other two homologous materials. With a lower concentration of mobile ions than in the β`` phase, and mobile ions separated by structural bottlenecks at aBR sites, the repulsion between neighboring $K^+$ ions becomes insufficient to drive fast conduction in K β-alumina as it does in K β``-alumina. By contrast, silver is effectively the smallest mobile ion due to its ability to adopt a two-coordinate configuration. The apparent snug fit of an Ag between oxygens enables faster conduction in Ag β-alumina, but is a detriment in Ag β``-alumina: the weak $Ag^+$-$Ag^+$ repulsion becomes insufficient to drive fast hopping. In both families of materials, the Na compound fits between the Ag and K systems in these trends.

The second factor is the identity and location of the charge-compensating dopant. We demonstrate that the Coulomb attraction of the $Mg_{Al}$ defects determines site-specific energetics and kinetics for mobile-ion diffusion in β``-aluminas (Figures 4 and 5). Furthermore, the defect locations anchor the mobile-ion vacancy ordering at low temperatures (Supplementary Note 2). In that case, unless the material is quenched and defect locations randomized, the highest-energy mobile-ion sites fit into a superlattice.

All this occurs in β``-aluminas despite the defects residing relatively far away from the conduction planes. While the $Al_2O_3$ host lattice remains rigid, it shields the mobile ions from long-duration trapping at charged defects when the defect is hidden inside the spinel block. The magnitudes of site-specific free energy differences are tens of meV in β``-aluminas (Figure 4e-g), but hundreds of meV in in β-aluminas (Figure S6e-g). Consequently, the trapping of mobile ions by oxygen interstitials in β-aluminas is stronger (Supplementary Note 4). Our simulation of Mg-substituted β-alumina (Supplementary Note 5) is consistent with this result: overall ionic conductivity is larger than for interstitial-based charge compensation. These results offer a simple guideline introducing charge-compensating aliovalent dopants into other solid electrolytes: static charge can be defused via covalency and physical separation. For smaller ions that are susceptible to strong binding by defects, such as Li, the incorporation of aliovalent dopants into covalently bonded frameworks or polyanions should mitigate their static charge (the "inductive effect")[45] and flatten the free-energy landscape for mobile ions.

Extended Data Figures

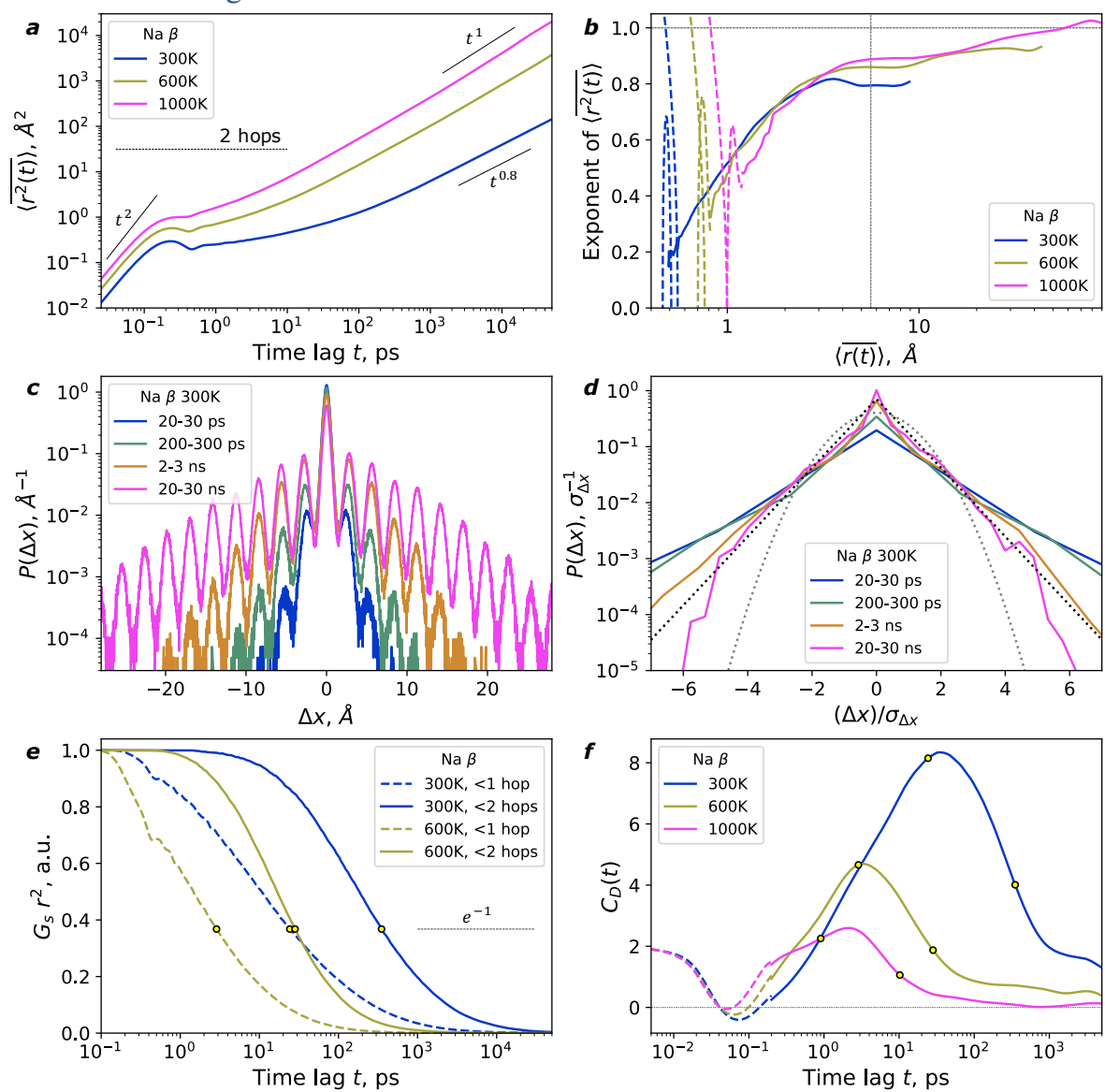

**Extended Data Figure 1 | Ion diffusion in Na β-alumina. a**, tMSD of mobile ions. **b**, exponent of tMSD vs time lag, plotted against the time-averaged displacement. The exponent does not reach unity until displacements are ≥ distances between defects. In **b**, the horizontal guide is the Fickian limit $t^1$, and the vertical line is one unit cell (5.6 Å, 2 hops). **c**, Time slices of the distribution of ion displacements Δx along [100] at 300 K. **d**, Distributions of ion displacements Δx along [100], each rescaled by its variance $\sigma_{\Delta x}$. Laplace and Gaussian distributions are shown as black and grey dotted lines, respectively. **e**, The probability of an ion remaining within 1.7 Å (<1 hop, dashed) or 4.6 Å (<2 hops, solid) of an initial position. The relative change in the timescale due to varying the distance cutoff by 0.1 Å is ≤10%. **f**, Diffusion kernel correlation $C_D$. $C_D$ peaks between one-hop and two-hop relaxation times at all simulated temperatures. In **b** and **f**, short-time checks of the ballistic limits (dashed), tMSD ∝ $t^2$ and $C_D \to 2$, respectively, used 100-ps trajectories recorded every 1 fs.

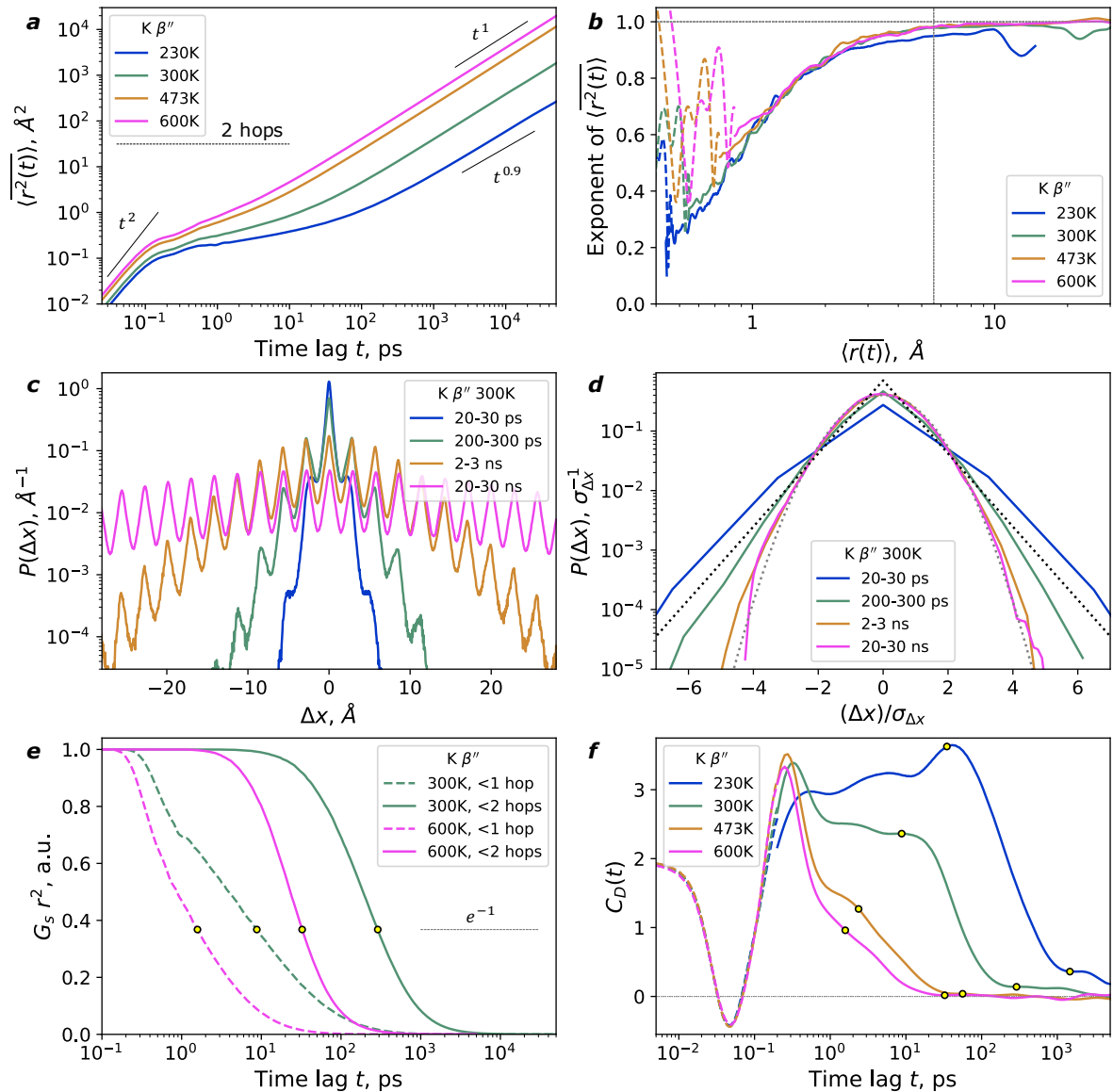

**Extended Data Figure 2 | Ion diffusion in K β''-alumina. a**, tMSD of mobile ions. **b**, exponent of tMSD vs. time lag, plotted against the time-averaged displacement. In **b**, the horizontal guide is the Fickian limit $t^1$, and the vertical line is one unit cell (5.6 Å, 2 hops). **c**, Time slices of the distribution of ion displacements $\Delta x$ along [100] at 300 K. **d**, Distributions of ion displacements $\Delta x$ along [100], each rescaled by its variance $\sigma_{\Delta x}$. Laplace and Gaussian distributions are shown as black and grey dotted lines, respectively. **e**, The probability of an ion remaining within 1.7 Å (<1 hop, dashed) or 4.6 Å (<2 hops, solid) of an initial position. The relative change in the timescale due to varying the distance cutoff by 0.1 Å is ≤10%. **f**, Diffusion kernel correlation $C_D$. Notably, unlike in Na β''-alumina (Figure 2), $C_D$ has multiple peaks. In **b** and **f**, short-time checks of the ballistic limits (dashed), tMSD $\propto t^2$ and $C_D \to 2$, respectively, used 100-ps trajectories recorded every 1 fs.

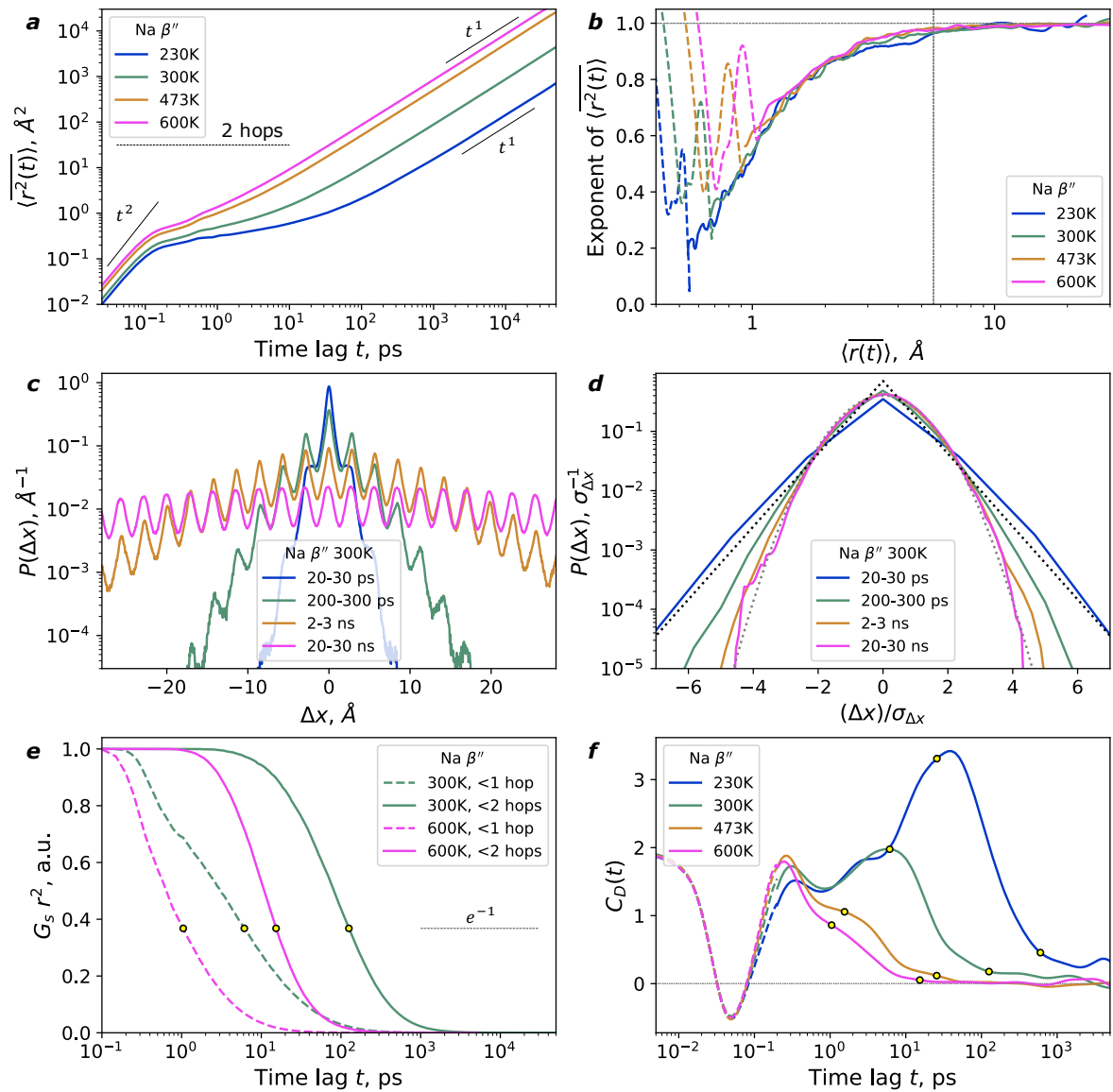

**Extended Data Figure 3 | Ion diffusion in Na β''-alumina with a quenched distribution of Mg$_{Al}$ defects. a**, tMSD of mobile ions. **b**, exponent of tMSD vs. time lag, plotted against the time-averaged displacement. In **b**, the horizontal guide is the Fickian limit $t^1$, and the vertical line is one unit cell (5.6 Å, 2 hops). The diffusion is Fickian (**a,b**) by contrast with Na β''-alumina without quenching (Figure 2). **c**, Time slices of the distribution of ion displacements Δ$x$ along [100] at 300 K. **d**, Distributions of ion displacements Δ$x$ along [100], each rescaled by its variance σ$_{Δx}$. Laplace and Gaussian distributions are shown as black and grey dotted lines, respectively. **e**, The probability of an ion remaining within 1.7 Å (<1 hop, dashed) or 4.6 Å (<2 hops, solid) of an initial position. The relative change in the timescale due to varying the distance cutoff by 0.1 Å is ≤10%. **f**, Diffusion kernel correlation $C_D$. Unlike the non-Gaussian parameter, $C_D$ has multiple peaks at 300 K. In **b** and **f**, short-time checks of the ballistic limits (dashed), tMSD ∝ $t^2$ and $C_D → 2$, respectively, used 100-ps trajectories recorded every 1 fs.

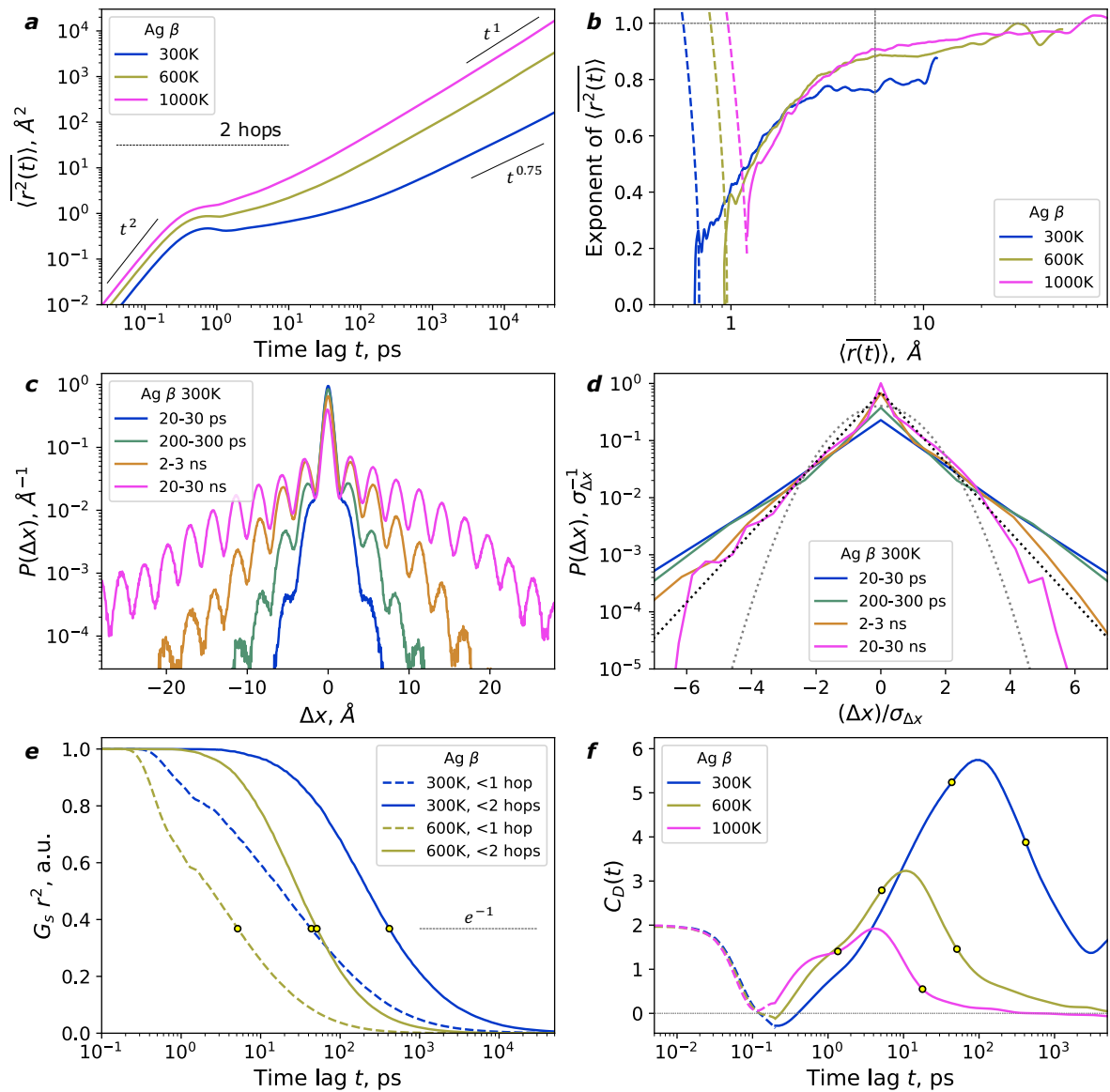

**Extended Data Figure 4 | Ion diffusion in Ag β-alumina. a**, tMSD of mobile ions. **b**, exponent of tMSD vs. time lag, plotted against the time-averaged displacement. As for Na β-alumina, the exponent does not reach unity until displacements are ≥ distances between defects. In **b**, the horizontal guide is the Fickian limit $t^1$, and the vertical line is one unit cell (5.6 Å, 2 hops). **c**, Time slices of the distribution of ion displacements $\Delta x$ along [100] at 300 K. **d**, Distributions of ion displacements $\Delta x$ along [100], each rescaled by its variance $\sigma_{\Delta x}$. Laplace and Gaussian distributions are shown as black and grey dotted lines, respectively. **e**, The probability of an ion remaining within 1.7 Å (<1 hop, dashed) or 4.6 Å (<2 hops, solid) of an initial position. The relative change in the timescale due to varying the distance cutoff by 0.1 Å is ≤10%. **f**, Diffusion kernel correlation $C_D$. As for Na β-alumina, $C_D$ peaks between one-hop and two-hop relaxation times at all simulated temperatures. In **b** and **f**, short-time checks of the ballistic limits (dashed), tMSD ∝ $t^2$ and $C_D \to 2$, respectively, used 100-ps trajectories recorded every 1 fs.

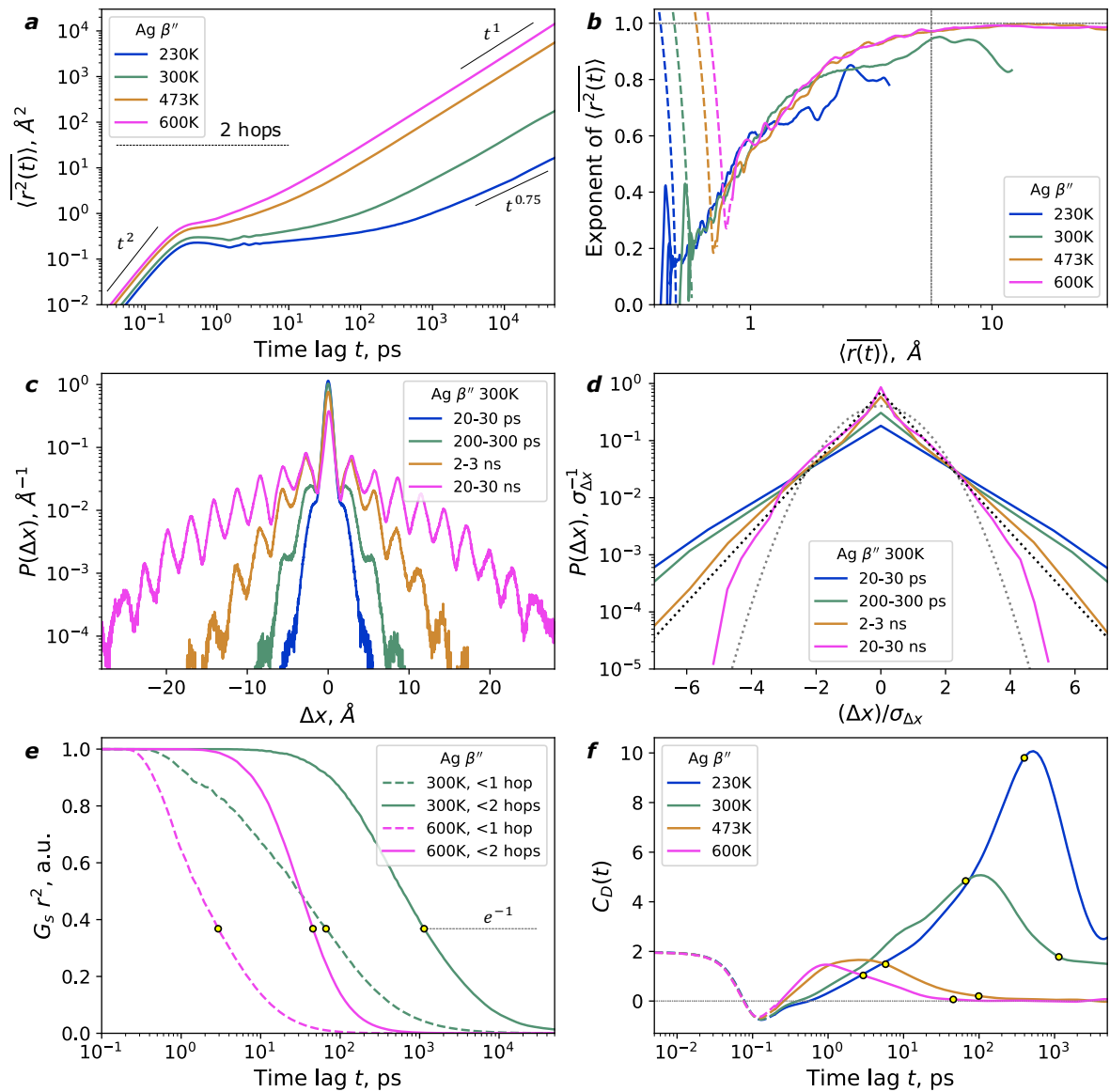

**Extended Data Figure 5 | Ion diffusion in Ag β''-alumina. a**, tMSD of mobile ions. **b**, exponent of tMSD vs. time lag, plotted against the time-averaged displacement. In **b**, the horizontal guide is the Fickian limit $t^1$, and the vertical line is one unit cell (5.6 Å, 2 hops). **c**, Time slices of the distribution of ion displacements $\Delta x$ along [100] at 300 K. **d**, Distributions of ion displacements $\Delta x$ along [100], each rescaled by its variance $\sigma_{\Delta x}$. Laplace and Gaussian distributions are shown as black and grey dotted lines, respectively. **e**, The probability of an ion remaining within 1.7 Å (<1 hop, dashed) or 4.6 Å (<2 hops, solid) of an initial position. The relative change in the timescale due to varying the distance cutoff by 0.1 Å is ≤10%. **f**, Diffusion kernel correlation $C_D$. In **b** and **f**, short-time checks of the ballistic limits (dashed), tMSD $\propto t^2$ and $C_D \to 2$, respectively, used 100-ps trajectories recorded every 1 fs.

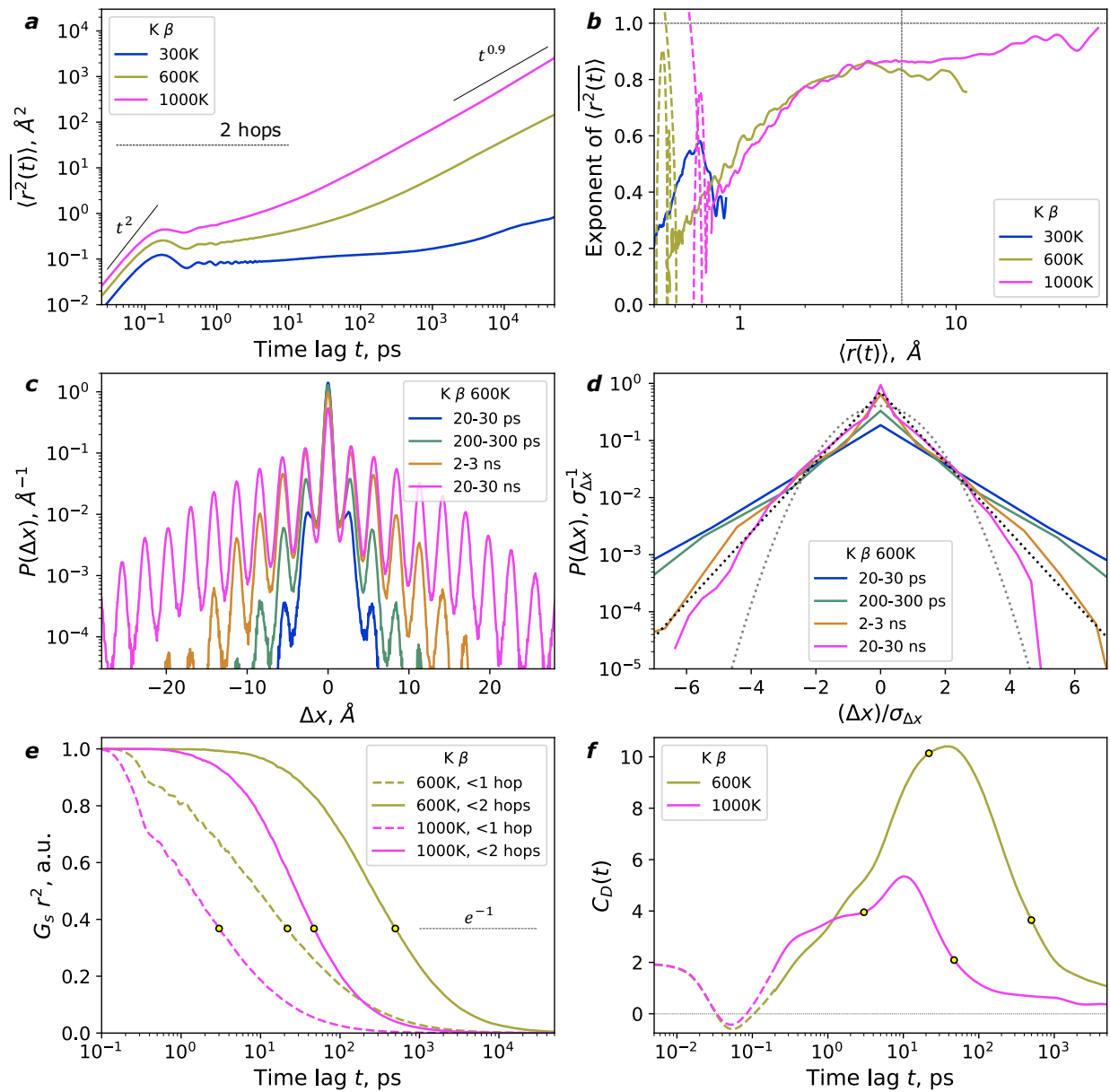

**Extended Data Figure 6 | Ion diffusion in K β-alumina. a**, tMSD of mobile ions. **b**, exponent of tMSD vs. time lag, plotted against the time-averaged displacement. As for Na β-alumina, the exponent does not reach unity until displacements are ≥ distances between defects. In **b**, the horizontal guide is the Fickian limit $t^1$, and the vertical line is one unit cell (5.6 Å, 2 hops). **c**, Time slices of the distribution of ion displacements $\Delta x$ along [100] at 300 K. **d**, Distributions of ion displacements $\Delta x$ along [100], each rescaled by its variance $\sigma_{\Delta x}$. Laplace and Gaussian distributions are shown as black and grey dotted lines, respectively. **e**, The probability of an ion remaining within 1.7 Å (<1 hop, dashed) or 4.6 Å (<2 hops, solid) of an initial position. The relative change in the timescale due to varying the distance cutoff by 0.1 Å is ≤10%. **f**, Diffusion kernel correlation $C_D$. In **b** and **f**, short-time checks of the ballistic limits (dashed), tMSD ∝ $t^2$ and $C_D \to 2$, respectively, used 100-ps trajectories recorded every 1 fs.

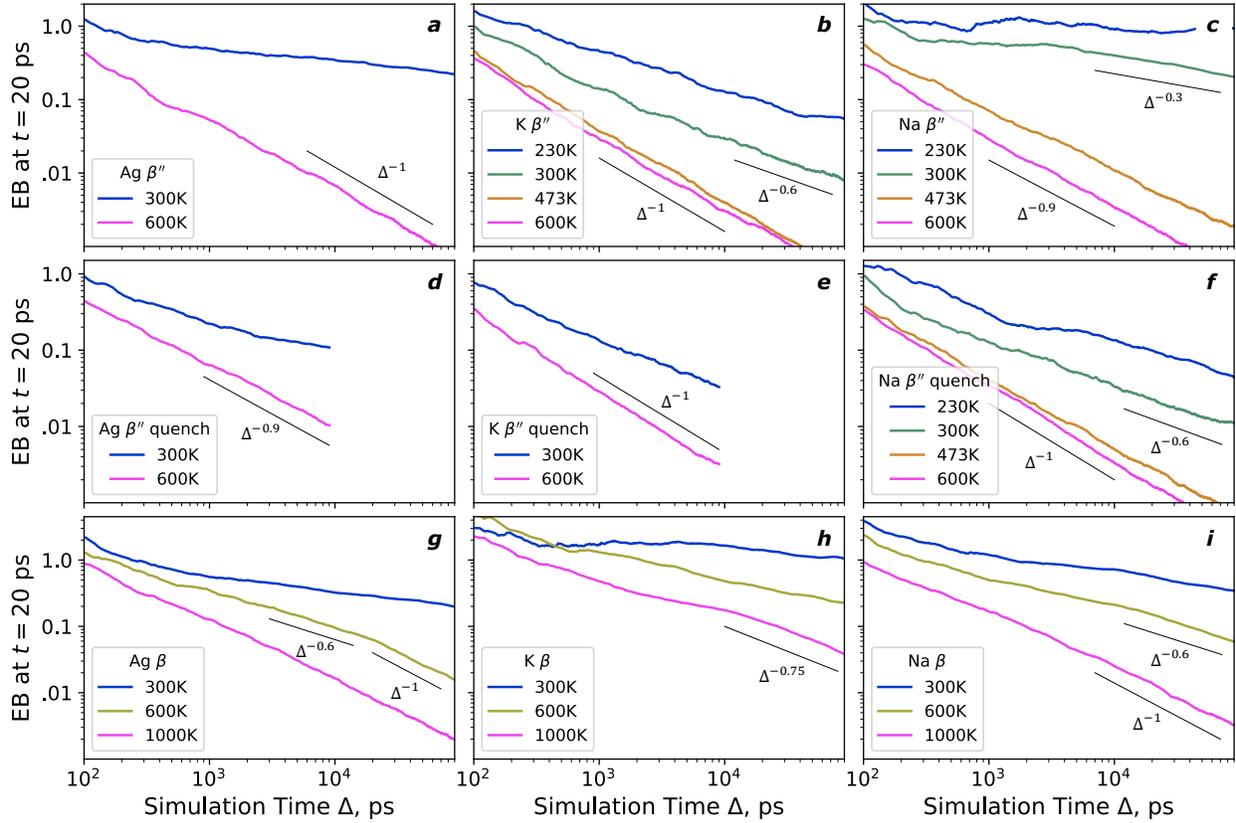

**Extended Data Figure 7 | Ergodicity breaking parameter (EB) versus simulation length Δ. a-c**, β``-aluminas. **d-f**, β``-aluminas with a simulated distribution of defects corresponding to quenching. **g-i**, β-aluminas. The time lag used in all cases was $t = 20$ ps, but the asymptotic dependences at long simulation lengths are not sensitive to the precise value of $t$ for $\Delta \gg t$. The noted power-law relations are guides to the eye, not quantitative fits. In all simulations where EB ∝ $\Delta^{-1}$, $C_D \to 0$ precedes it. This is seen most clearly in Ag β-alumina at 600 K (**g**), where $C_D \to 0$ for $t \approx 10$ ns (Extended Data Figure 4), and EB ∝ $\Delta^{-1}$ starting at $\Delta \approx 20$ ns.

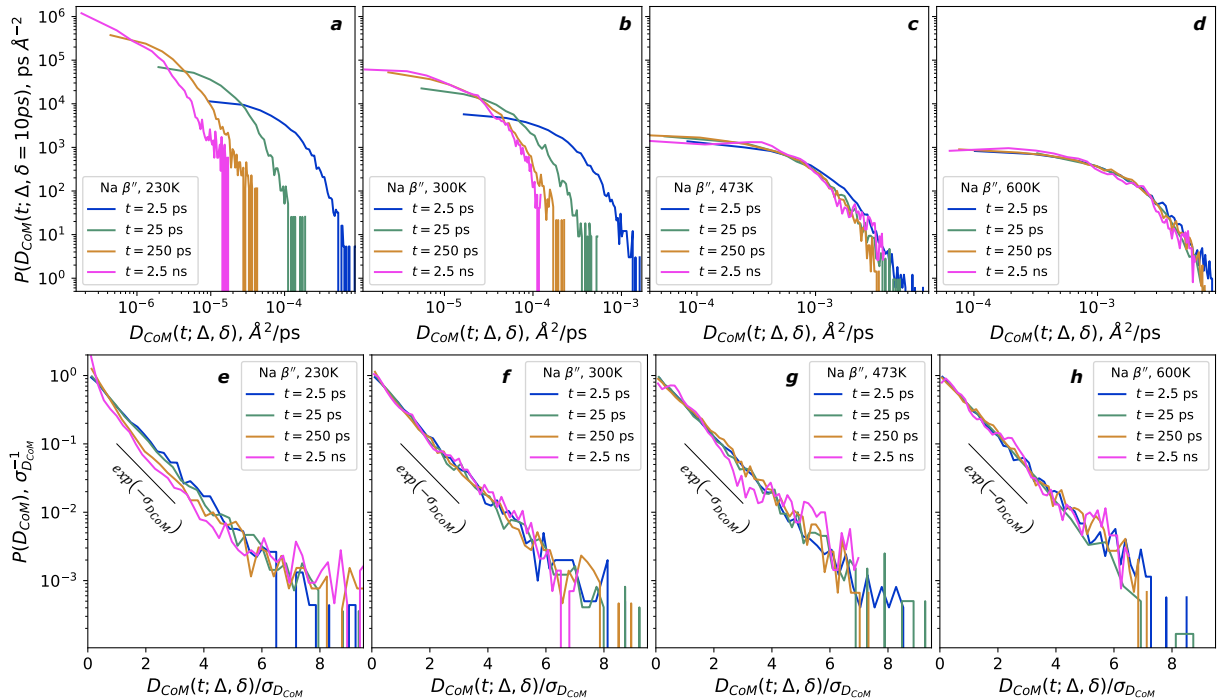

**Extended Data Figure 8 | Distributions of the center-of-mass diffusion coefficient $D_{CoM}$ in Na β''-alumina. a-d**, absolute values, **e-h**, rescaled by the standard error at each time lag. (**a,e**) 230 K, (**b,f**) 300 K, (**c,g**) 473 K, (**d,h**) 600 K. The rescaled distributions (**e**)-(**h**) are exponential. Notably, at 230 K, the distribution becomes wider rather than narrower with increasing time lag, suggesting the possibility of further glass-like collective dynamics at long timescales. At 230 K, the ordering of the mobile ions extends across the entire simulation cell within each conduction plane (Figure S2).

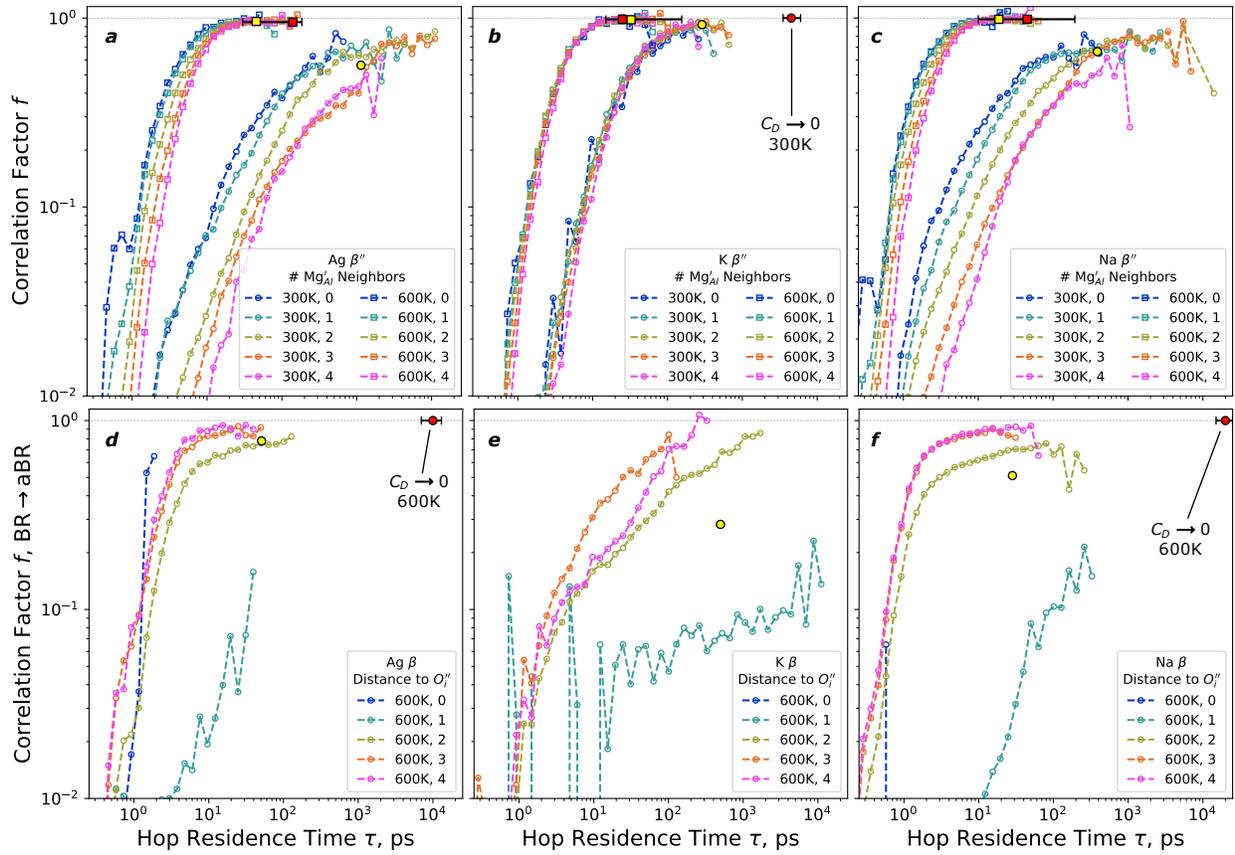

**Extended Data Figure 9 | Correlation factor $f$ for hops, dis-aggregated by location relative to defects, versus hop residence times. a-c**, β''-aluminas. **d-f**, β-aluminas. For β-aluminas, hops starting in the low-energy Beevers-Ross sites are considered here. The sites with the most neighboring defects (for β'') or those closest to defects (for β) have the lowest correlation factor for a given residence time. Yellow symbols: two-hop relaxation time from $G_s$. Black horizontal ranges: $C_D \to 0$ at 600 K. At 300 K, $C_D \to 0$ only for K β''-alumina.